\documentclass[preprint,12pt]{elsarticle}

\usepackage{epsfig}
\usepackage{amssymb}
\usepackage{amsmath}
\usepackage{subcaption}
\usepackage{float}
\usepackage{url}

\usepackage[breaklinks]{hyperref}
\usepackage{multirow}
\usepackage{graphicx}

\journal{}

\begin{document}

\begin{frontmatter}
\title{SIR Simulation of COVID-19 Pandemic in Malaysia: Will the Vaccination Program be Effective?}
\author[label1]{W. K. Wong}
\author[label1]{Filbert H. Juwono\corref{cor1}}
\author[label2]{Tock H. Chua}

\address[label1]{Department of Electrical and Computer Engineering, Faculty of Engineering and Science, Curtin University Malaysia, 98009 Miri, Sarawak, Malaysia}
\address[label2]{Department of Pathobiology and Medical Diagnostics, Faculty of Medicine and Health Sciences, University Malaysia Sabah, 88400 Kota Kinabalu, Malaysia}

\cortext[cor1]{Corresponding author}

\begin{abstract}
Since the end of 2019, COVID-19 has significantly affected the lives of people around the world. Towards the end of 2020, several COVID-19 vaccine candidates with relatively high efficacy have been reported in the final phase of clinical trials. Vaccines have been considered as critical tools for opening up social and economic activities, thereby lessening the impact of this disease on the society. This paper presents a simulation of COVID-19 spread using modified Susceptible-Infected-Removed (SIR) model under vaccine intervention in several localities of Malaysia, i.e. those cities or states with high relatively COVID-19 cases such as Kuala Lumpur, Penang, Sabah, and Sarawak. The results show that at different vaccine efficacy levels (0.75, 0.85, and 0.95), the curves of active infection vary slightly, indicating that vaccines with efficacy above 0.75 would produce the herd immunity required to level the curves. In addition, disparity is significant between implementing or not implementing a vaccination program. Simulation results also show that lowering the reproduction number, $R_0$ is necessary to keep the infection curve flat despite vaccination. This is due to the assumption that vaccination is mostly carried out gradually at the assumed fixed rate. The statement is based on our simulation results with two values of $R_0$: 1.1 and 1.2, indicative of reduction of $R_0$ by social distancing. The lower $R_0$ shows a smaller peak amplitude about half the value simulated with $R_0=1.2$. In conclusion, the simulation model suggests a two-pronged strategy to combat the COVID-19 pandemic in Malaysia: vaccination and compliance with standard operating procedure issued by the World Health Organization (e.g. social distancing).
\end{abstract}


\end{frontmatter}

\section{Introduction}
Coronavirus disease 2019 (COVID-19) is caused by the 2019 novel coronavirus (2019-nCOV) first reported in the city of Wuhan, China. The epidemic began in December 2019, when several adults in Wuhan presented with serious pneumonia \cite{singhal20}. Since then, the infection has spread globally and the number of cases increased exponentially, with more than 90 million people infected by January 2021 \cite{worldometer}. The COVID-19 pandemic impacted the socio-economic wellbeing for many countries in 2020 \cite{pak20}. It has been believed that the vaccination would be the most promising alternative to get life back to normal. To date, there have been at least 166 vaccine candidates undertaking pre-clinical and clinical trials \cite{jeyanathan20}.

In the fourth quarter of 2020, several pharmacological companies reported high efficacy rates of their vaccine candidates \cite{cnbc}. By mid-December 2020, 57 vaccine candidates were in clinical research, including 40 candidates in Phase I–II trials and 17 candidates in Phase II–III trials. In Phase III trials, several COVID-19 vaccines demonstrated efficacy as high as 95\% in preventing symptomatic COVID-19 infections \cite{polack20}. Various levels of efficacy had been reported for the advance phases of clinical trials: Sinopharm (79\%) \cite{straittimes}, Pfizer BioNTech (95\%) \cite{polack20}, Moderna (94.5\%) \cite{time} at advance phases of clinical trials. A survey was conducted in \cite{lazarus20} to gauge the public acceptance of COVID-19 vaccines in 19 countries. From the results, 71.5\% of respondents reported that they would be very or very likely to take vaccines and 61.4\% reported accepting employer’s recommendation to do so. Acceptance rate discrepancies ranged from almost 90\% (China) to less than 55\% (Russia).

With the proposed vaccination campaign, countries are considering to open economic operations to reactivate the socio-economy activities. In Malaysia, vaccines were scheduled to arrive in the mid of first quarter of 2021 \cite{thestar}. However, according to a public survey, approximately one-third of Malaysian people were still unsure of the safety of COVID-19 vaccines \cite{malaymail2}. Hence, it is crucial to create public awareness on the importance of vaccination.

In this paper, we we present the results of simulating the effectiveness of vaccination in reducing the infection rates in Malaysia. In particular, we focus on Kuala Lumpur, the capital of Malaysia and three states, i.e. Penang, Sabah, and Sarawak. Further, we apply a modified Susceptible-Infected-Removed (SIR) model to incorporate constant vaccination rate based on the Malaysia Government's planning. SIR has been widely used to model COVID-19 pandemic spread \cite{COOPER20,SHAROV20}. Note that mathematical modelling has been a long progressive field due to its inherent importance to health policy makers. This will help to answer some fundamental questions, such as the effectiveness of the vaccination program given various efficacy rates and current developments. However, it needs to be highlighted that these projections are only based on the proposed model. Furthermore, the modified SIR model does not take into account the stochasticity of the imported cases. This research will serve as a projection to observe the effects of vaccination at different efficacy rates and to show the importance of vaccination.

\section{Previous Work}
In \cite{Bertozzi16732}, the authors discussed various models for forecasting the spread of COVID-19 including the common compartmental models (SIR and Susceptible–Exposed–Infected–Removed (SEIR)), exponential growth model, and self exiting branch model. Despite there are various methods, the principle remains the same i.e. to find some parameters which enable the model to represent the actual recorded data. Obviously, compartmental models have been highly popular in modeling COVID-19 as seen from massive research works in this area. SIR model was used to fit the COVID-19 data and then used to forecast the number of cases in Senegal \cite{Balde20}. In Italy, particle swarm optimization (PSO), a form of stochastic optimization method, was used to fit SEIR COVID-19 model to actual data \cite{Godio_2020}. Acu\~na-Zegarra, et. al. \cite{acuna20} developed an optimal control problem to design vaccination strategies with various efficacy rates to gauge the COVID-19 pandemic situation. It was difficult to gauge the efficiency of the vaccine to provide immunity to the receivers as the clinical trials are based on procedures and evaluation based on the efficacy. SIR models have been developed for some countries  (China, South Korea, India, Australia, and USA) \cite{COOPER20}. In \cite{SHAROV20}, SIR model was created to analyze the effectiveness using the data from 15 European countries.

In general, most pandemic related models are an extension of the basic SIR model proposed in \cite{kermack27}. This model is built on the premise that entire population can be divided into three states: susceptible (S), infected (I), and removed (R) states. The infected group continuously infects the population until isolation is performed. The assumption requires the recovered population to receive 100\% immunity. The SIR model follows a deterministic model with linear infection rate. In \cite{koufi19}, the authors modified the existing SIR to be stochastic model and included vertical transmission and vaccination and non-linear incidence and vaccination. Current vaccination SIR models mostly consider newborn vaccination as a result of previous and on-going efforts. However, the principles could be similarly applied in similar manner with slight modifications.

\section{Mathematical Models}
\subsection{SIR Model}
The basic SIR model without isolation and vaccination effect is given by the following expressions
\begin{equation}
    \label{eqn_S}
    \frac{dS(t)}{dt} = \frac{-\beta I(t)S(t)}{N},
\end{equation}
\begin{equation}
    \label{eqn_I}
    \frac{dI(t)}{dt} = \frac{\beta I(t)S(t)}{N} - \gamma I(t),
\end{equation}
\begin{equation}
    \label{eqn_R}
    \frac{dR(t)}{dt} = \gamma I(t),
\end{equation}
\begin{equation}
    \label{eqn_N}
    N = S(t)+I(t)+R(t), \forall t,
\end{equation}
where $S(t)$ represents the number of people in an area at time $t$ who are susceptible to the disease infection and can be infected by the infectious people, $I(t)$ represents the number of people in an area at time $t$ who infected and infectious due to the spread of the virus, $R(t)$ represents the number of people in an area at time $t$ who are removed from the infected state, $N$ is the total number of people in the area at time $t$, $\beta$ is a constant showing infectivity rate, i.e. expected number of people infected by an infectious person, and $\gamma$ is a constant showing removal rate, i.e. expected number of people removed from the infected state. The ratio of $\beta$ and $\gamma$ is called as reproduction number, i.e. $R_0 = \beta/\gamma$. Reproduction number shows the average number of secondary infections coming from an infected person assuming everyone is in susceptible state.

In practice, the constant $\beta$ depends on how the society practice hygiene and social distancing adhering to the standard operating procedure and the constant $\gamma$ is related to the people who have either recovered or deceased. The initial number of susceptible people, $S(t=0)$ can be found using the data released by the Government. From \eqref{eqn_N}, we have $S(t=0)=N-R(t=0)-I(t=0)$. We can approximate $I(t=0)$ from the identification rate, $p$ and the average ratio of asymptomatic and symptomatic cases, $\varphi$. For Malaysia, we assume a fixed ratio of asymptomatic and symptomatic cases, i.e. $\varphi = 0.7$ \cite{thestar2}. Let $X(t)$ be the number of new daily cases identified. The approximate number of initial infected cases is $I(t=0)=X(t=0)/p \times 0.7$.

\subsection{Modified SIR Model}
\begin{figure}
        \centering
        \includegraphics[width=5.0in]{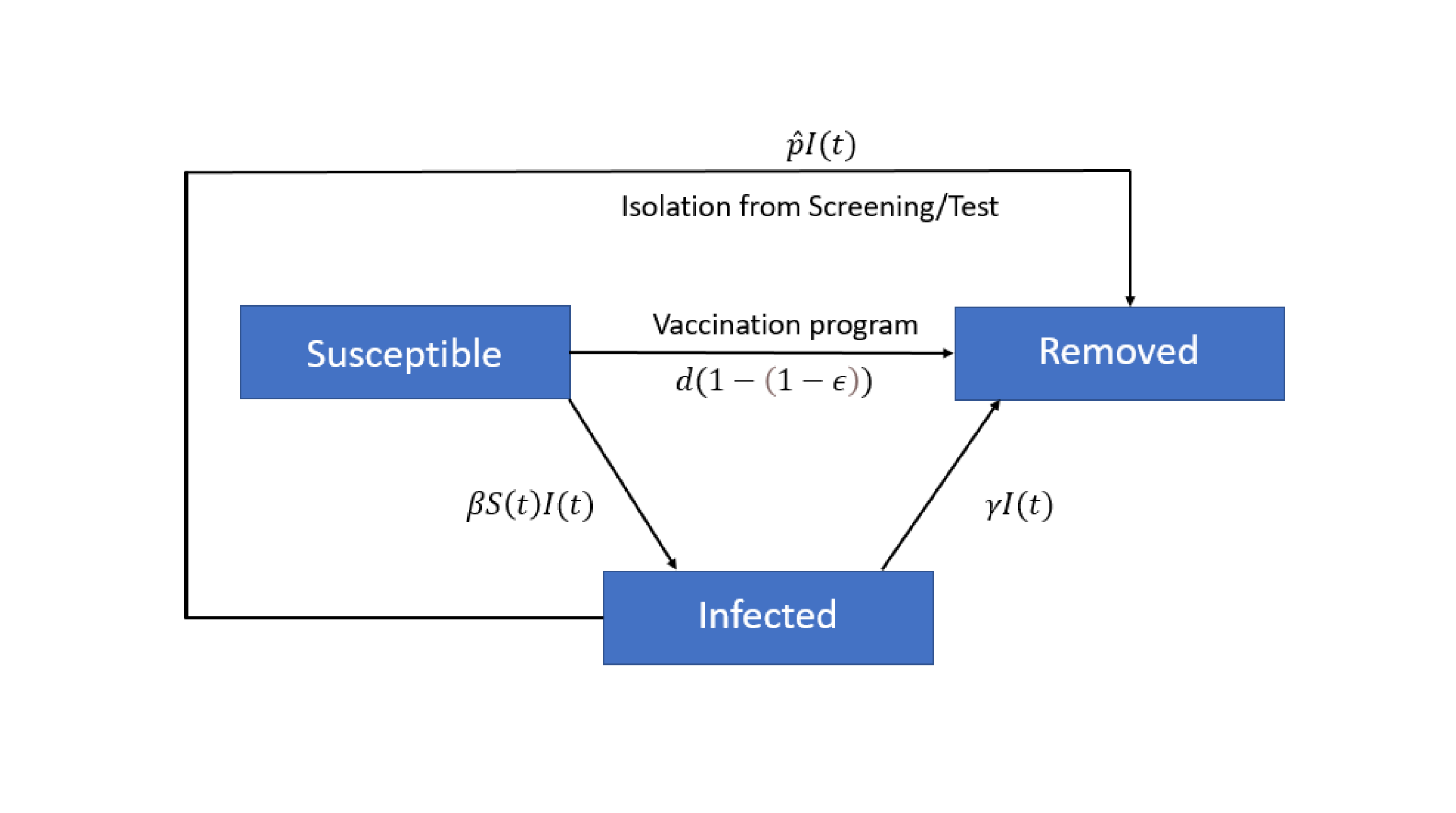}
        \caption{State diagram of modified SIR model with vaccination strategy.}
        \label{fig_states}
    \end{figure}

Fig. \ref{fig_states} depicts the proposed state diagram. We introduce a new term, vaccination rate, $d$ to the existing SIR model as shown in the figure. The vaccination rate is defined as the ratio of target population and completion time. This assumption is based on the demographic and logistical challenges of administering the vaccines.

According to the state diagram and using the model from \cite{YONGZHEN11}, we modify the SIR model for the $S(t)$ and $R(t)$ expressions to be
    \begin{equation}
    \label{eqn_modS}
        \frac{dS(t)}{dt}=-d-\frac{\beta I(t)S(t)}{N}, \text{ s.t. } S(t)>0
    \end{equation}
    
    \begin{equation}
    \label{eqn_modR}
        \frac{dR(t)}{dt}=d+\gamma I(t).
    \end{equation}
Note that we consider only the constant rate vaccination and ignore the death rate (removed state is equivalent to recovered state) as the number of deaths is small compared to the total population. In \cite{YONGZHEN11}, the dynamics of the vaccination is based on number of S(t). However, we know that vaccination program will be progressive and based on a fixed rate on the remaining susceptible class. We may assume that $d$ is equivalent to an ersatz efficiency in vaccination since no information on this is attained at the moment globally.

Further, we take into account the vaccine efficacy factor in the modified $S(t)$ and $R(t)$ formulae. We introduce a multiplier factor $\epsilon$ which is defined as the ratio of the number of positive cases in the vaccinated samples and the number of positive cases in the placebo samples during clinical trials. Therefore, the vaccine efficacy is given by $(1-\epsilon)$. Expressions \eqref{eqn_modS} and \eqref{eqn_modR} considering $\epsilon$ factor are given by
    \begin{equation}
    \label{eqn_modSep}
        \frac{dS(t)}{dt}=-d \epsilon-\frac{\beta I(t)S(t)}{N}, \text{ s.t. } S(t)>0
    \end{equation}
    
    \begin{equation}
    \label{eqn_modRep}
        \frac{dR(t)}{dt}=d (1-\epsilon)+\gamma I(t).
    \end{equation}

In case that self-isolation from random checks or contact tracing is conducted everyday, again, we need to re-modify the $I(t)$ and $R(t)$ expressions in \eqref{eqn_I} and \eqref{eqn_modRep}, respectively. Finally, we have the following SIR model which incorporates vaccination and isolation strategies
    \begin{equation}
    \label{eqn_modSepiso}
        \frac{dS(t)}{dt}=-d \epsilon-\frac{\beta I(t)S(t)}{N}, \text{ s.t. } S(t)>0
    \end{equation}
    
    \begin{equation}
    \label{eqn_modIepiso}
        \frac{dI(t)}{dt}= \frac{\beta I(t) S(t)}{N}-\hat{\gamma}I(t),
    \end{equation}
    
    \begin{equation}
    \label{eqn_modRepiso}
        \frac{dR(t)}{dt}=d (1-\epsilon)+\hat{\gamma} I(t),
    \end{equation}
where $\hat{\gamma}=\gamma-\hat{p}$ and $\hat{p}$ is the isolation rate of positive infected individuals in the population. Note that the reproduction number is now $R_0 = \beta/\hat{\gamma}$.

\section{Simulation Setup}
\subsection{General Simulation Setup}
Table \ref{table_pop} shows the simulation details which include the location, total population, initial number of people in removed (recovered) state, $R(t)$, and initial number of new infected cases, $X(t)$. We set 1 January 2021 as $t = 0$. The data are obtained from Government website and other sources. We use two reproduction numbers for simulations, $R_0 = 1.1$ and $R_0 = 1.2$. The overall Malaysia $R_0$ fluctuates between 1 and 1.5. The latest $R_0$ projection has been estimated to be average of 1.1 until 31 May 2021 \cite{moh}. Moreover, we will use Runge-Kutta differential solver method to obtain the simulation curve.

\begin{table}
\centering
\caption{Details of simulation parameters.}
\begin{tabular}{cccc} 
\hline
\multicolumn{1}{c}{\textbf{Location}} & \multicolumn{1}{c}{\textbf{Population ($N$)}} & \multicolumn{1}{c}{\textbf{$R(t=0)$}} & \multicolumn{1}{c}{\textbf{$X(t=0)$}}  \\ 
\hline
Malaysia (overall)              & 32.7 million                      &120,000                      &2,500                            \\
Kuala Lumpur                 &  1.808 million                                  & 12,494     &   202                          \\
Penang                       &   1.767 million                                 &   4,160        & 60                             \\
Sabah                        &   3.54 million                                 &    36,074       &  186                           \\
Sarawak                      &   2.16 million                                 &     1,115          & 8                             \\
\hline
\end{tabular}
\label{table_pop}
\end{table}

\subsection{Malaysia Vaccination Program Details}
Table \ref{table_vacc} shows the vaccines delivery set by Malaysian Government. The table only shows the confirmed purchases inked by the government while many other negotiations are on-going. The government is also in final negotiations with China’s Sinovac for 14 million doses, CanSino Biologics for 3.5 million doses, and and Russia for 6.4 million shots of Sputnik V vaccine \cite{bloomberg}. However, it is clear that most vaccines will arrive only in the second and third quarter of 2021 onwards based on projections. In fact, the Malaysian Government has expressed the desire to cover 80\% of the population \cite{straittimes2}. In this paper, we assume that the vaccination rate is constant and that 75\% of the target population will be vaccinated within a year. The figure is reasonable as the inked deals have already covered almost 50\% of the Malaysian population (see the cumulative percentage in Table \ref{table_vacc}).

\begin{table}
\centering
\caption{COVID-19 vaccination plan in Malaysia.}
\begin{tabular}{cccc} 
\hline
\multicolumn{1}{c}{\textbf{Quarter (2021)}} & \multicolumn{1}{c}{\textbf{Doses in million}} & \multicolumn{1}{c}{\textbf{Vaccination }} & \multicolumn{1}{c}{\textbf{Cumulative}}  \\ 
\hline
1   & 1 (Pfizer) & 1.52\%    & 1.52\%   \\
\multirow{2}{*}{2}   & 8.1 (Pfizer) & 12.2\%    & \multirow{2}{*}{23.48\%}   \\
 & 6.4 (Covax) & 9.78\% & \\
 \multirow{2}{*}{3}   & 5.8 (Pfizer) & 8.86\%    & \multirow{2}{*}{42.12\%}   \\
 & 6.4 (Covax) & 9.78\% & \\
 4   & 4.3 (Pfizer) & 6.57\%    & 48.69\%   \\
\hline
\end{tabular}
\label{table_vacc}
\end{table}

Since most vaccines have reported efficacy more than 70\%, we will apply 75\%-95\% efficacy in our simulations. We assume 75\% inoculation by end $t=540$ with the various efficacy rates. This would yield 0.13\% vaccination rate per day (18 months to complete 75\% inoculation excluding those that have received natural immunity due to recovery). We can see that vaccines will be reserved for front liners and high risk populations in the first quarter and almost none for the public. Hence, our simulation will start from the second quarter of 2021. This is a proper figure to simulate the realistic scenario. If vaccination is performed at mid February 2021, the effective immunity will only be achieved by March 2021 (assume 15 days to reach immunity).

\section{Results and Discussion}
\subsection{Curve Fitting}
\begin{figure}[H]
    \centering
	\includegraphics[width=4in]{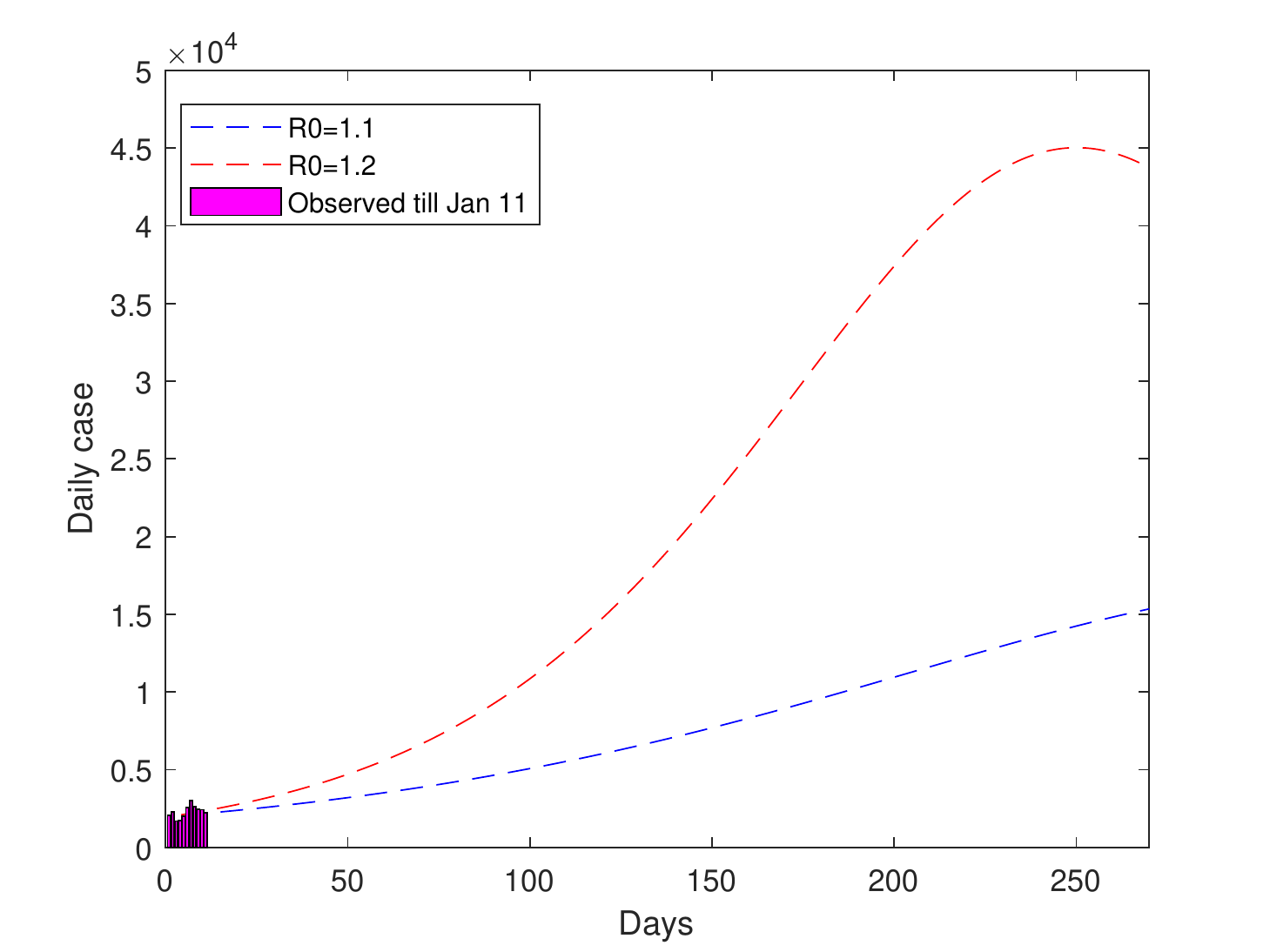}
	\caption{Projected daily case with $R_0 = 1.1$ and $R_0 = 1.2$.}
	\label{fig_daily_cases}
	\end{figure}

The optimized parameters, $\beta$ and $\hat{\gamma}$, in \eqref{eqn_modSepiso}, \eqref{eqn_modIepiso}, and \eqref{eqn_modRepiso} are acquired by curve-fitting the daily cases from 1 January 2021 to 11 January 2021 such that the reproduction numbers are $R_0 = 1.1$ and $R_0 = 1.2$, following the Government projection. The projected daily cases for 270 days is shown in Fig. \ref{fig_daily_cases}. Daily cases projection can be acquired by $\Delta R$ in the R compartment of the population. We acquire this best fit by systematically combining the parameters (grid search). Our best fit parameters give $\hat{\gamma}=0.08$ with $\beta=0.100$ and $\beta=0.108$ for $R_0=1.1$ and $R_0=1.2$, respectively. 

\begin{figure}[H]
	\centering
	\includegraphics[width=5in]{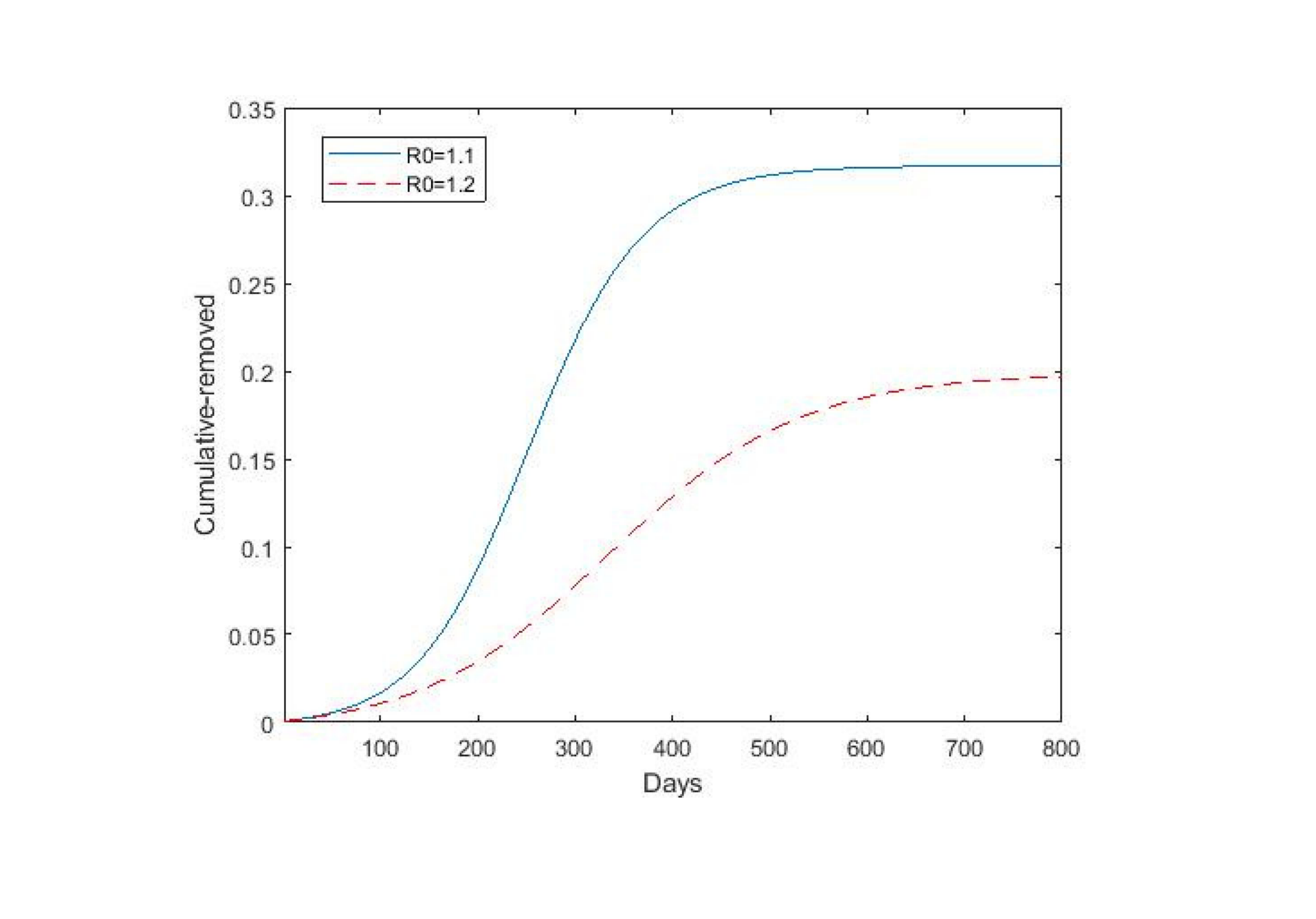}
	\caption{Removed (recovered) projections for 800 days with $R_0 = 1.1$ and $R_0 = 1.2$.}
	\label{fig_removed800}
	\end{figure}

Fig. \ref{fig_removed800} shows the removed cases projection without vaccination. It seems to suggest at $R_0=1.1$ and $R_0=1.2$, both cases will only cause natural immunity at 17\% and 31\%, respectively. This situation is based on the premise that no new imported cases have been added into the community. Note that herd immunity can only be reached if about 70\% of the population is immune to the virus. This is not recommended, of course, considering that Malaysia has a comparatively high elderly (people over 60 years of age) population, i.e. about 1.4 million.

\subsection{Simulation Results}
\begin{figure}[thb!]
	\begin{subfigure}{0.5\textwidth}
	\includegraphics[width=\linewidth]{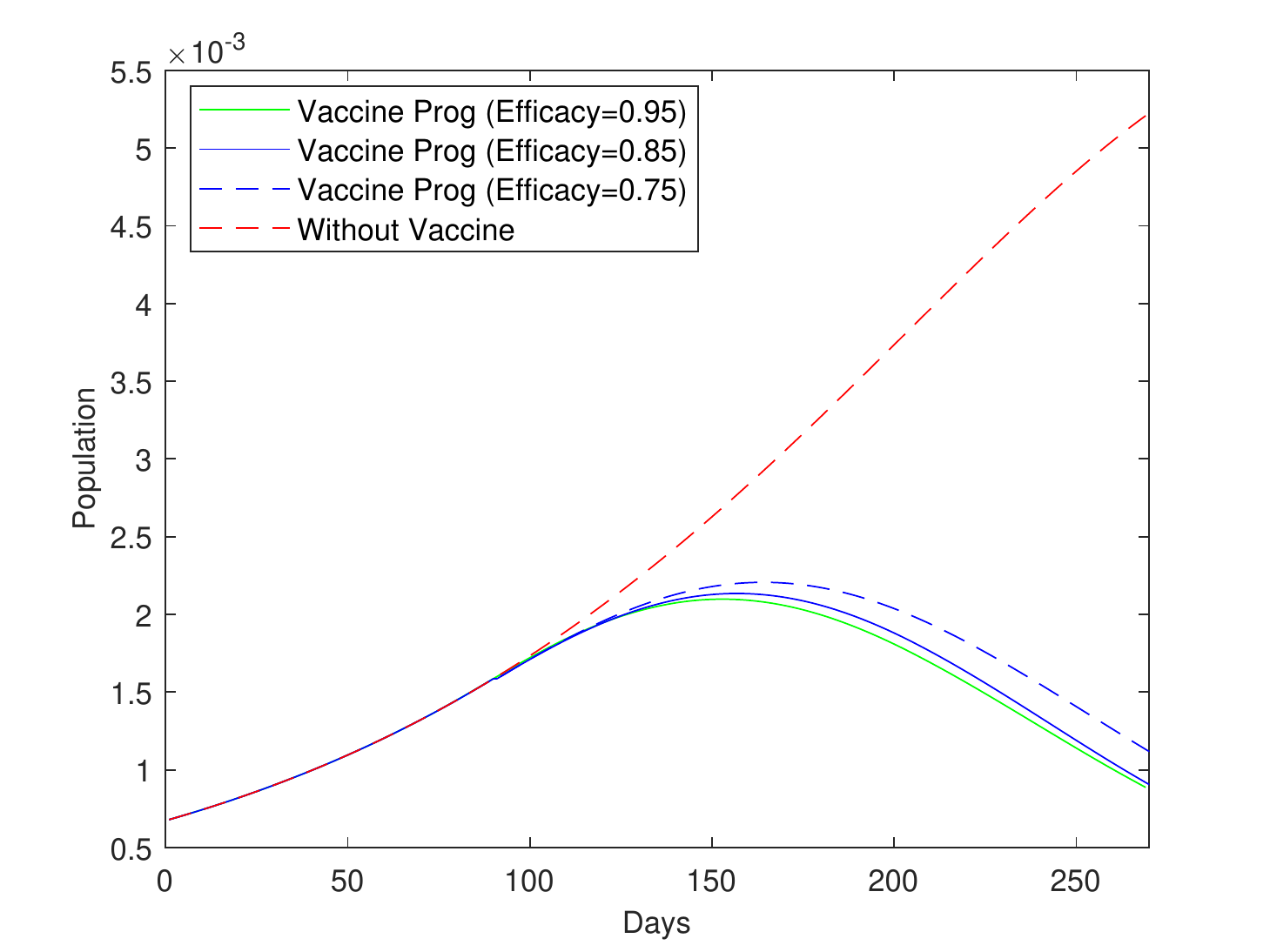}
	\caption{Malaysia.}
	\label{fig_malaysia11}
	\end{subfigure}
	\begin{subfigure}{0.5\textwidth}
	\includegraphics[width=\linewidth]{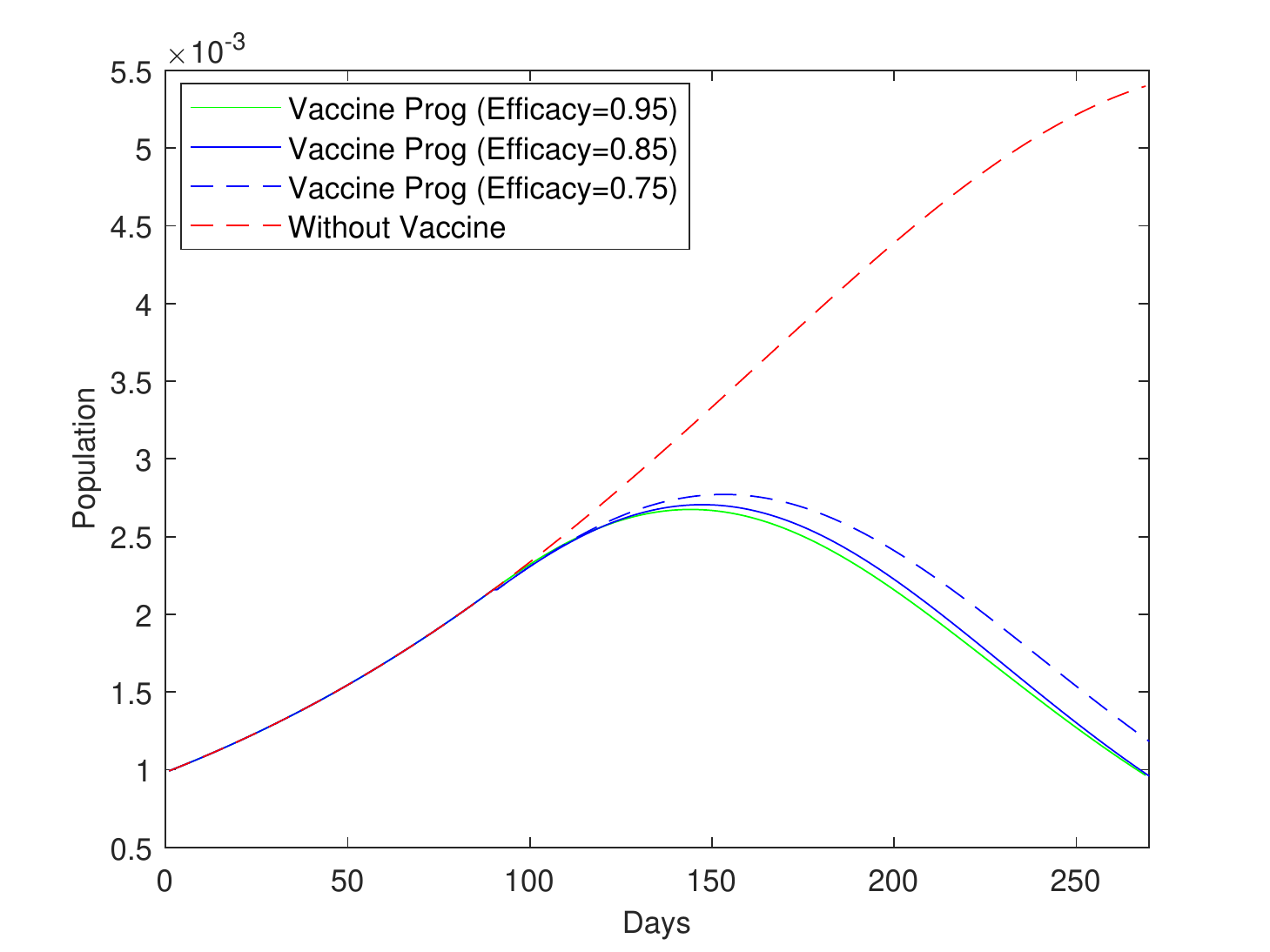}
	\caption{Kuala Lumpur.}
	\label{fig_kul11}
	\end{subfigure}
	\begin{subfigure}{0.5\textwidth}
	\includegraphics[width=\linewidth]{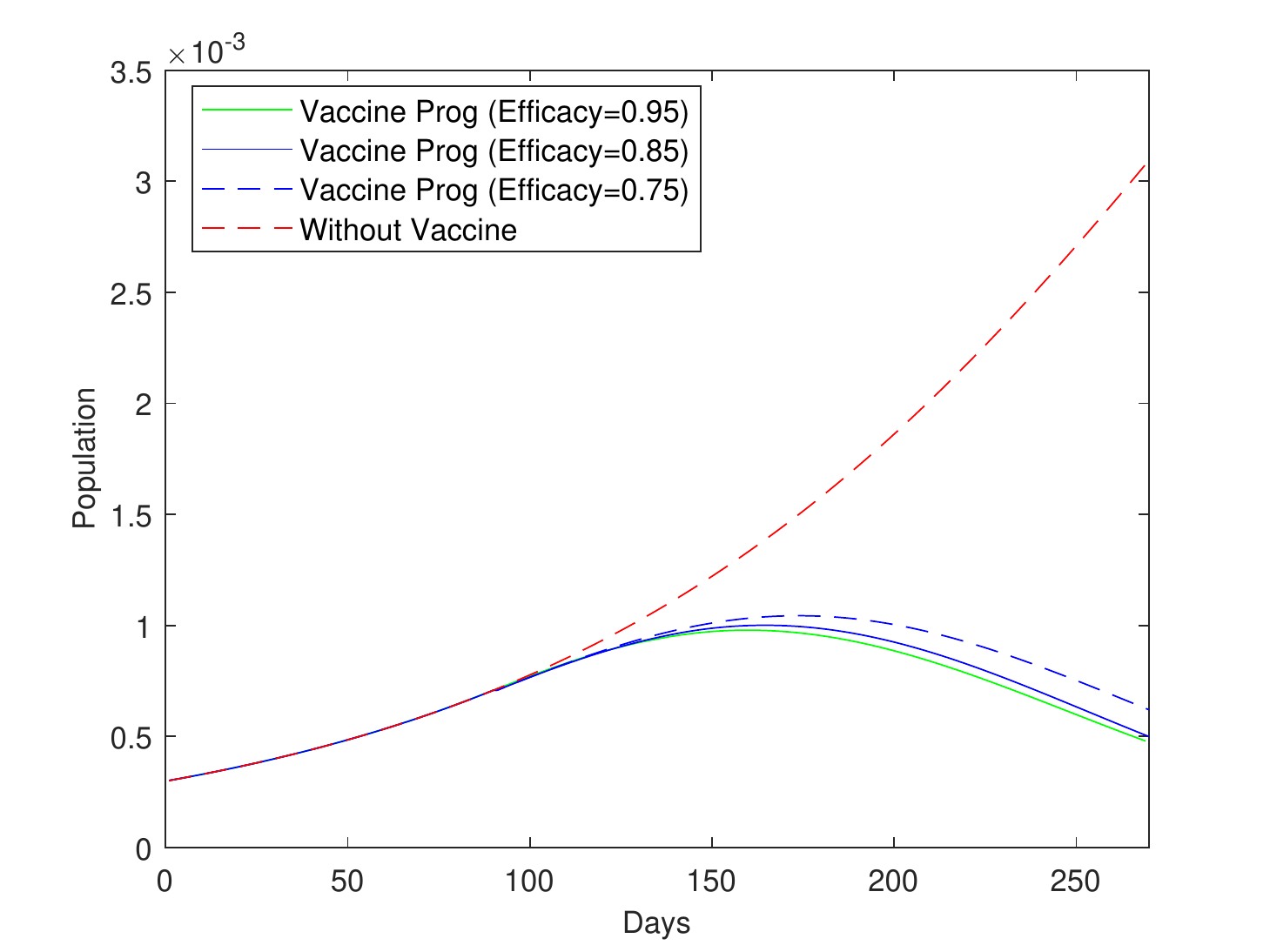}
	\caption{Penang.}
	\label{fig_penang11}
	\end{subfigure}
	\begin{subfigure}{0.5\textwidth}
	\includegraphics[width=\linewidth]{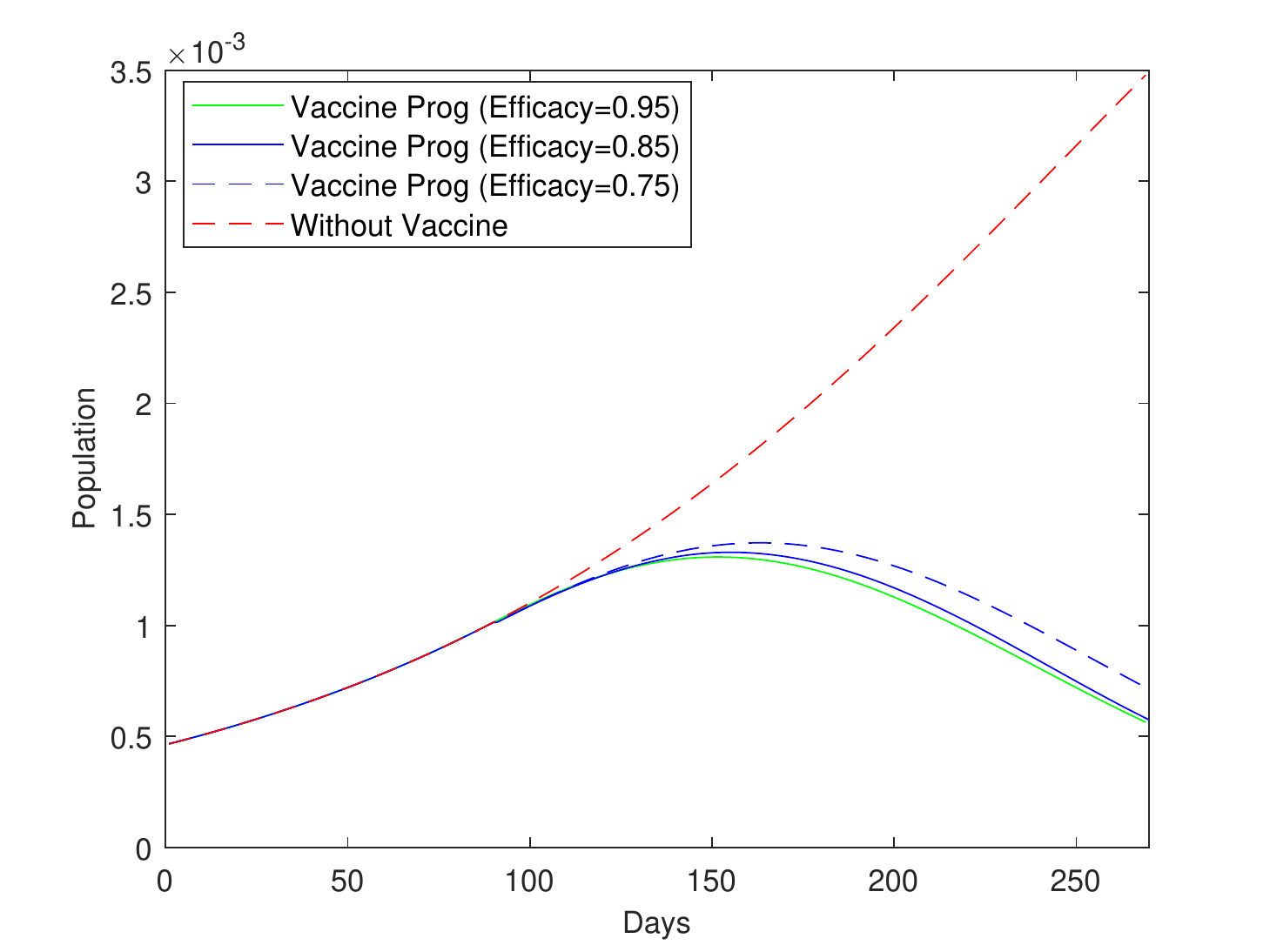}
	\caption{Sabah.}
	\label{fig_sabah11}
	\end{subfigure}
	\begin{subfigure}{0.5\textwidth}
	\includegraphics[width=\linewidth]{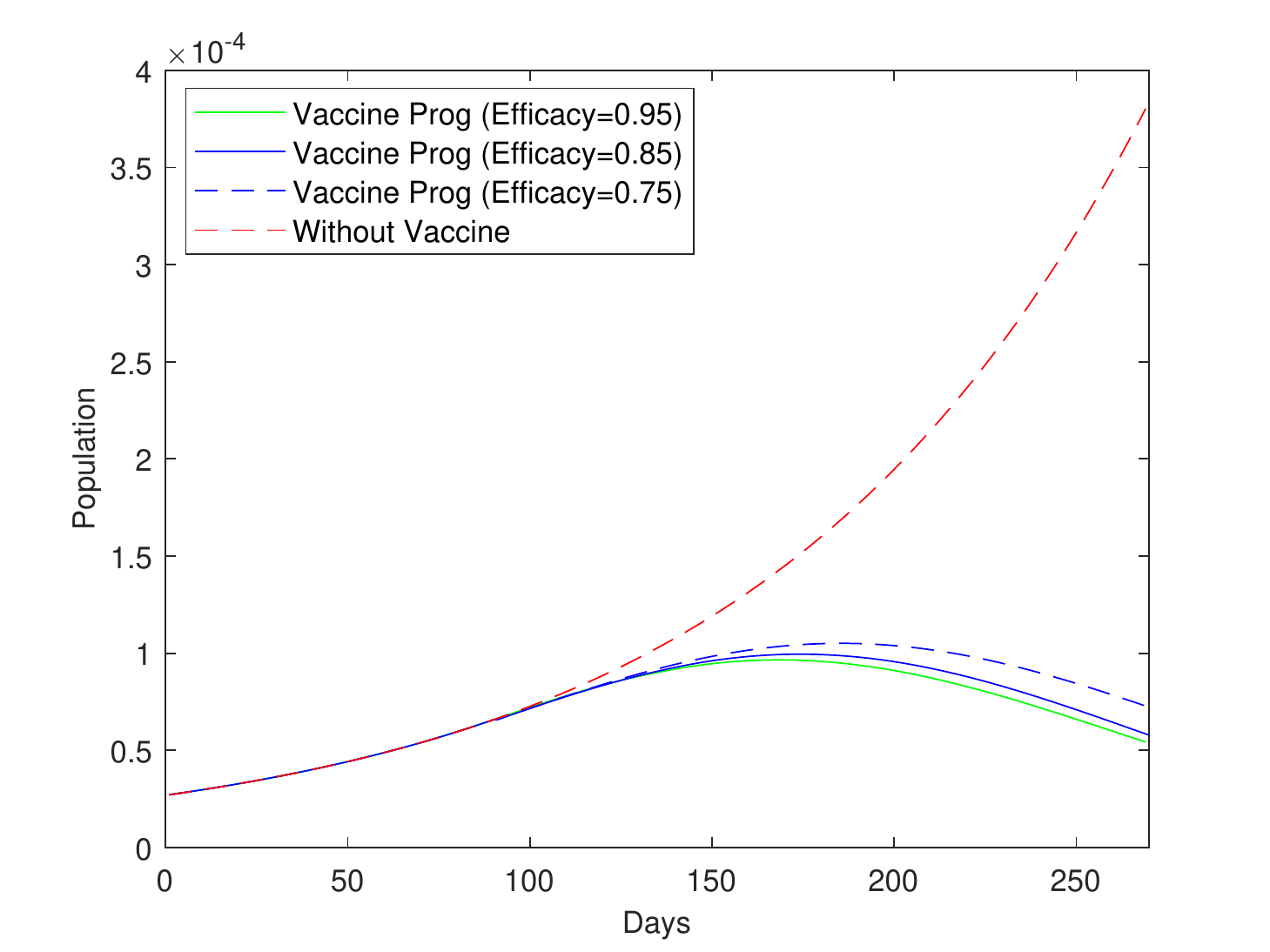}
	\caption{Sarawak.}
	\label{fig_sarawak11}
	\end{subfigure}
	\caption{Modified SIR simulation with and without vaccination for the various localities, $R_0=1.1$.}
	\label{fig_projection11}
	\end{figure}

\begin{figure}[thb!]
	\begin{subfigure}{0.5\textwidth}
	\includegraphics[width=\linewidth]{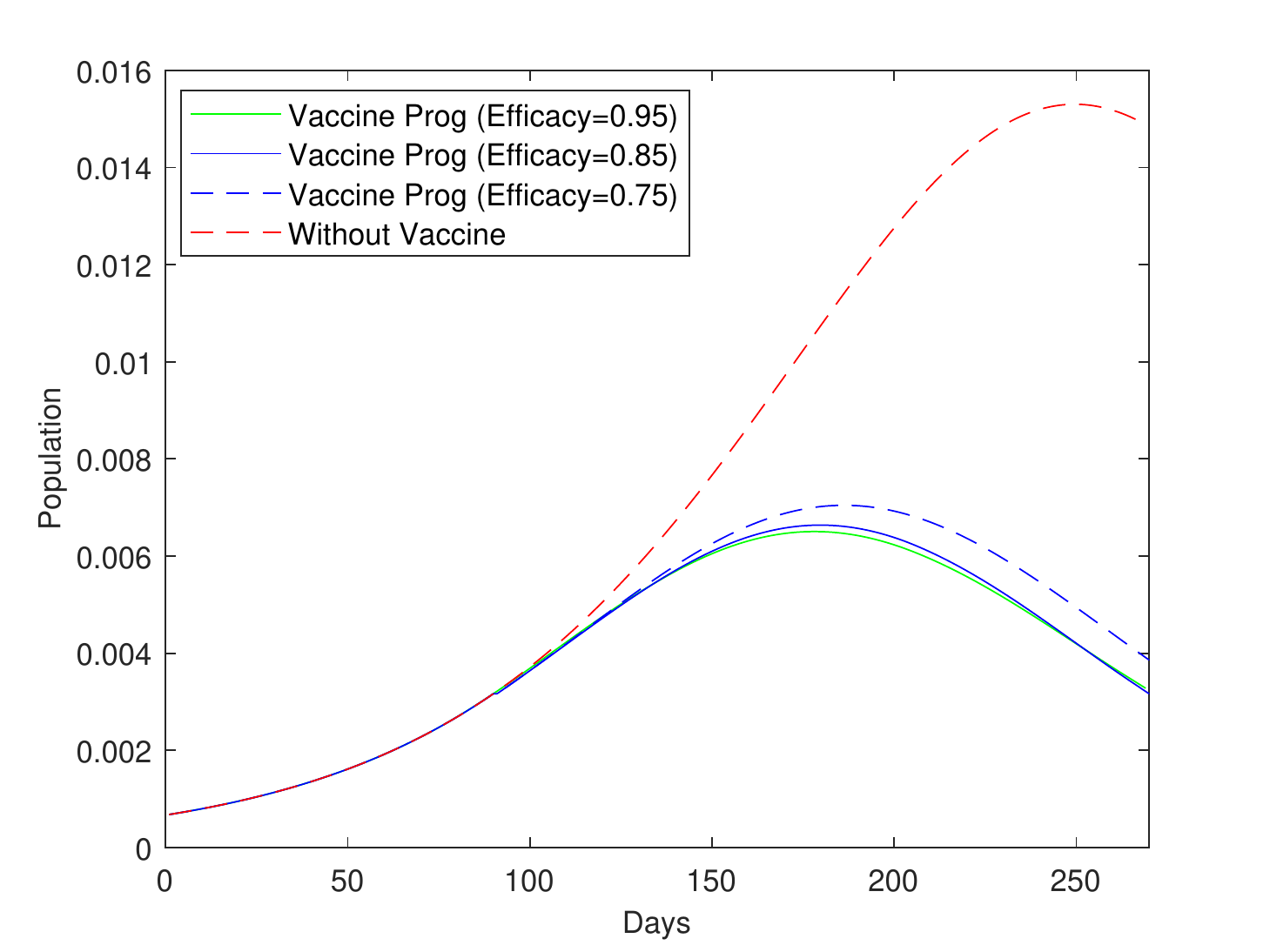}
	\caption{Malaysia.}
	\label{fig_malaysia12}
	\end{subfigure}
	\begin{subfigure}{0.5\textwidth}
	\includegraphics[width=\linewidth]{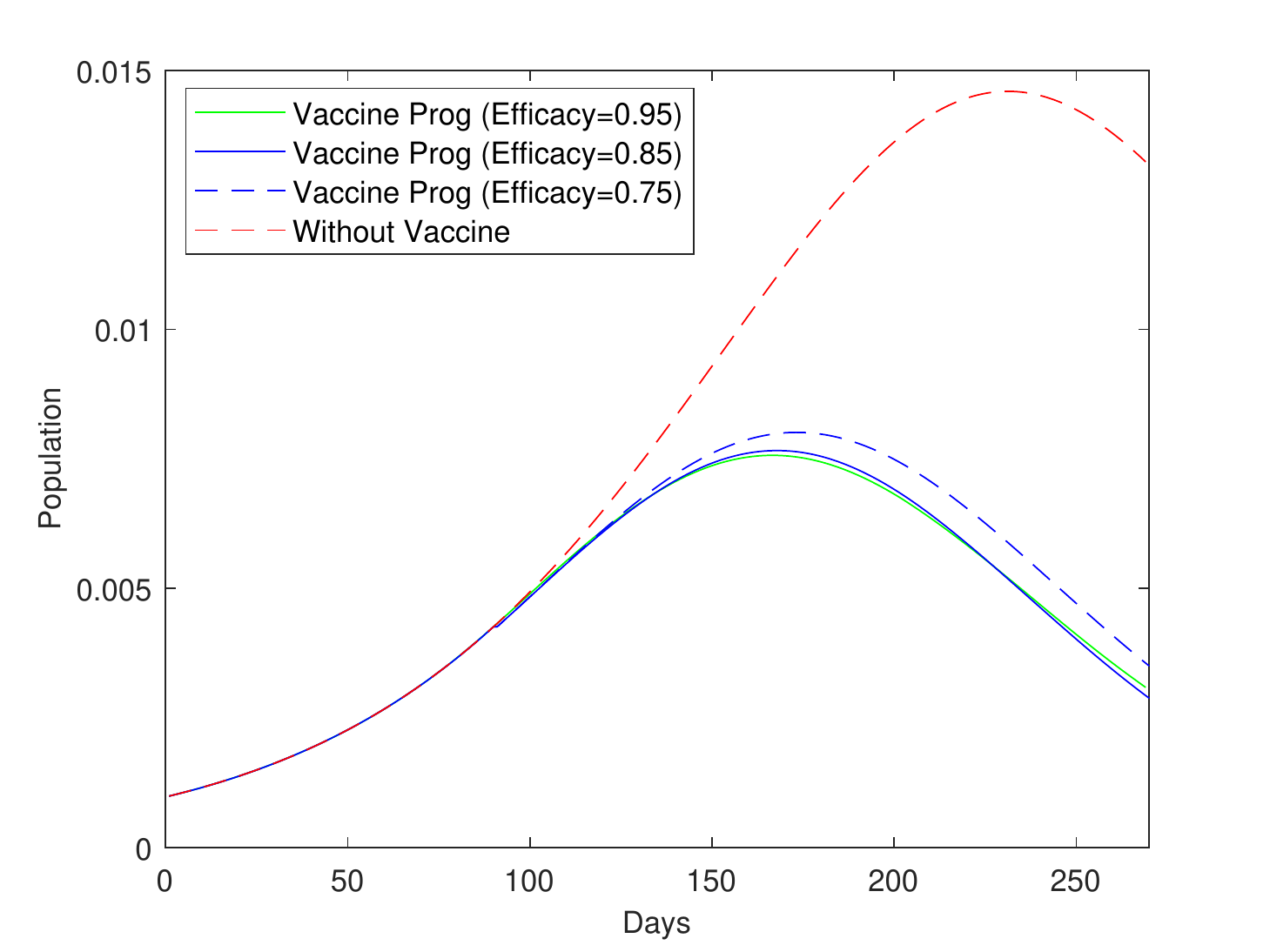}
	\caption{Kuala Lumpur.}
	\label{fig_kul12}
	\end{subfigure}
	\begin{subfigure}{0.5\textwidth}
	\includegraphics[width=\linewidth]{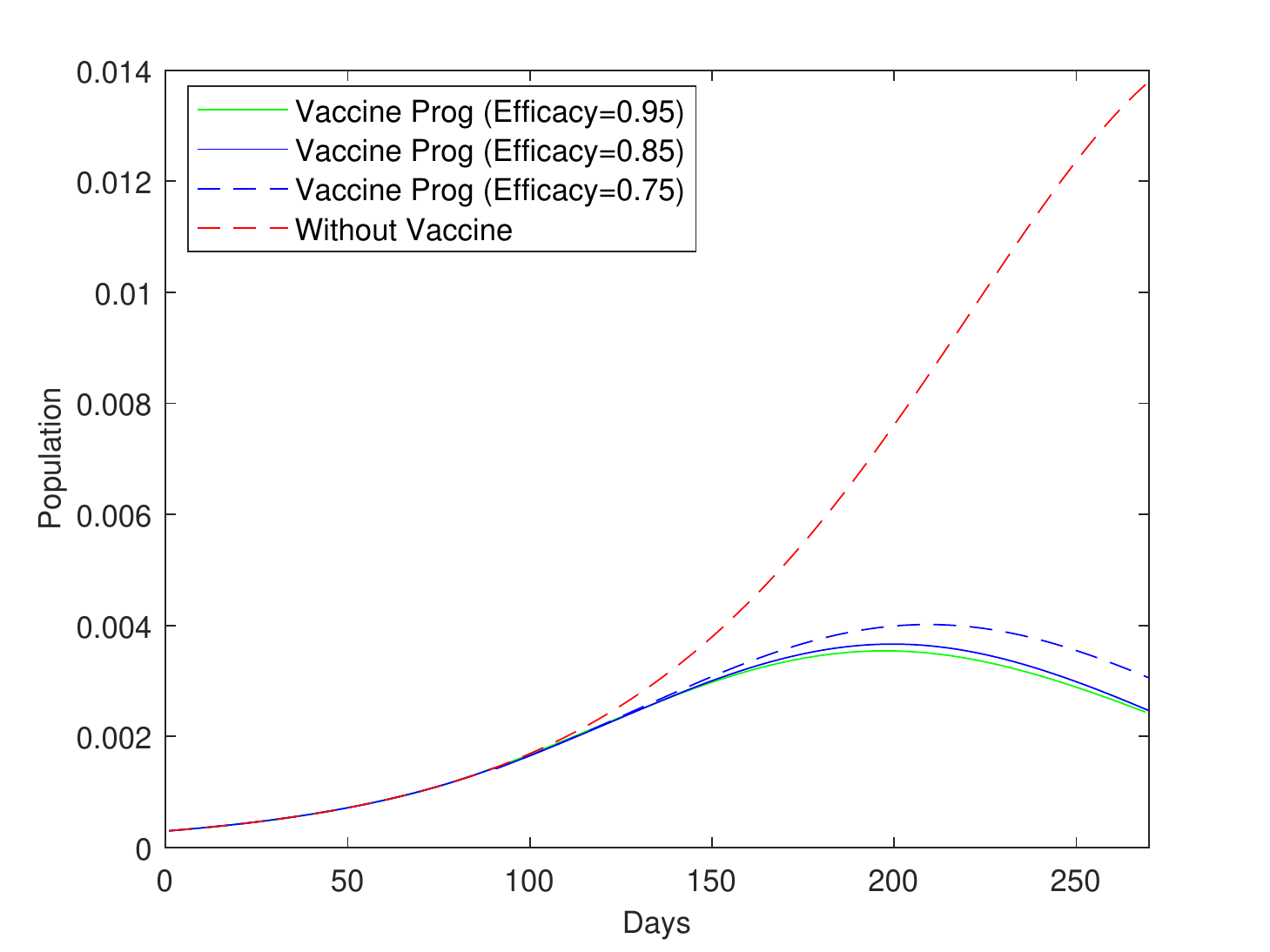}
	\caption{Penang.}
	\label{fig_penang12}
	\end{subfigure}
	\begin{subfigure}{0.5\textwidth}
	\includegraphics[width=\linewidth]{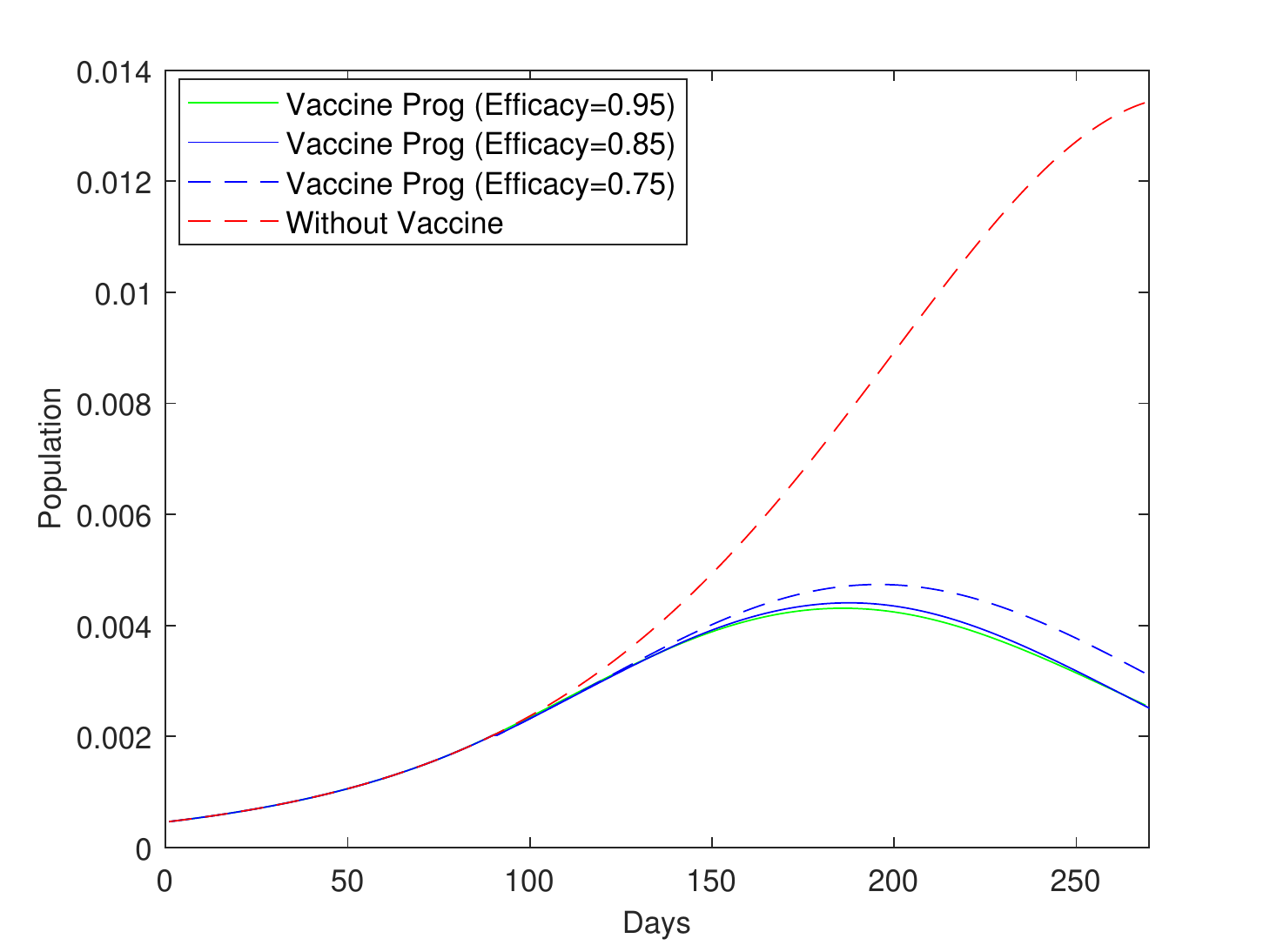}
	\caption{Sabah.}
	\label{fig_sabah12}
	\end{subfigure}
	\begin{subfigure}{0.5\textwidth}
	\includegraphics[width=\linewidth]{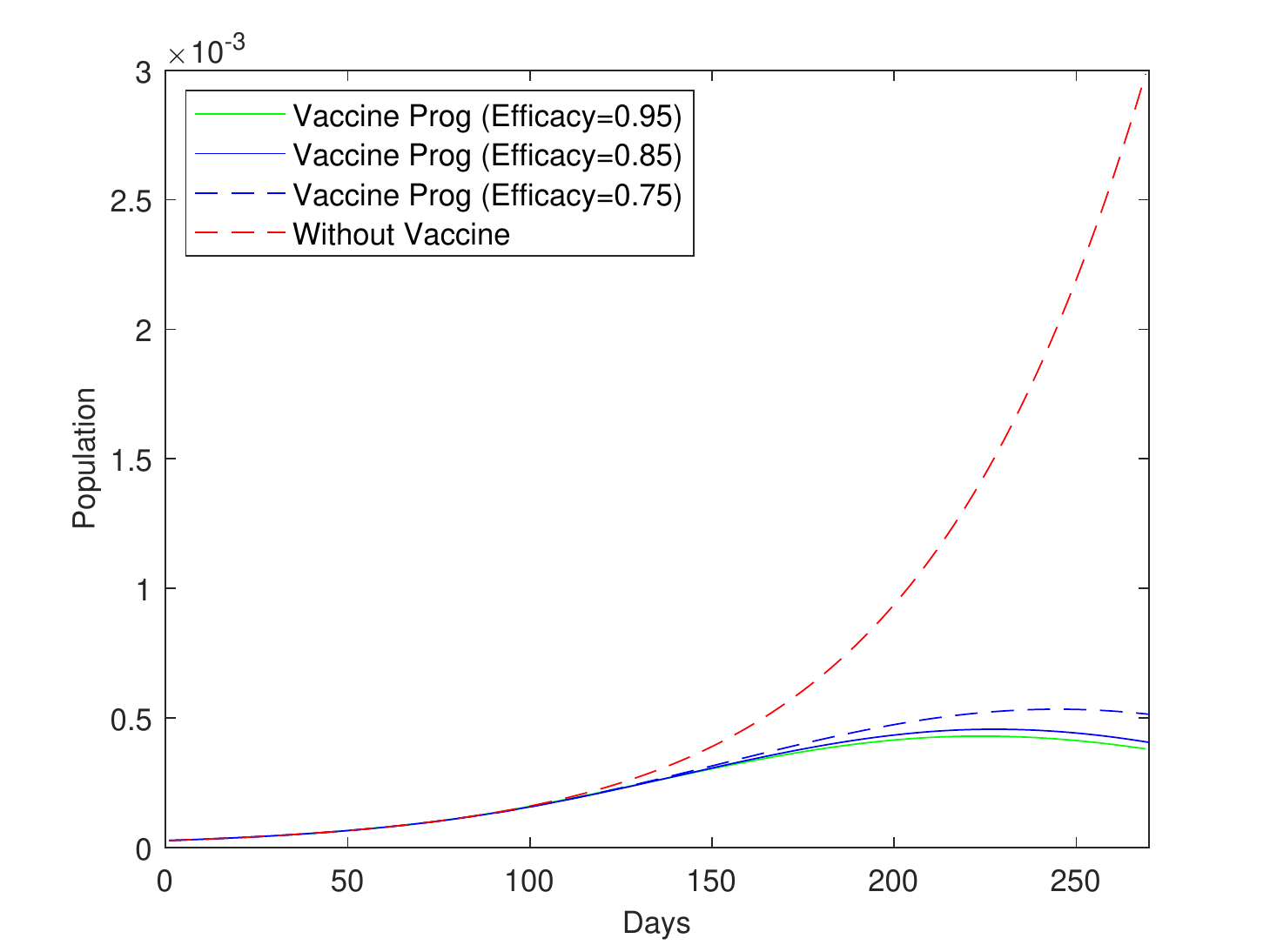}
	\caption{Sarawak.}
	\label{fig_sarawak12}
	\end{subfigure}
	\caption{Modified SIR simulation with and without vaccination for the various localities, $R_0=1.2$.}
	\label{fig_projection12}
	\end{figure}

The projection plots for 270 days using the given population and vaccination data and formulae \eqref{eqn_modSepiso}, \eqref{eqn_modIepiso}, and \eqref{eqn_modRepiso} with $R_0 = 1.1$ and $R_0 = 1.1$ are shown in Figs. \ref{fig_projection11} and \ref{fig_projection12}, respectively. We observe three efficacy rates: 0.75, 0.85, and 0.95. It is worth mentioning that we apply a normalized population i.e. $N=1.00$ and the individual state is scaled respectively.

In Fig. \ref{fig_projection11}, it can be seen that vaccination program is effective which is shown by the big decrease in the percentage of population getting infected. In other words, the curve gets flattened more rapidly at about 10 - 40 days after day 90 ($t=90$). Keep in mind that we assume that vaccination will take place in the second quarter of 2021. In Fig. \ref{fig_projection12}, we can observe similar trend except with a more protruding curve due to higher production number. Except for Sarawak, most of the cases show deviation of peak within 10-20 days. As observed, the amplitude (peak of curve) is also found to be lower even in early stages of vaccination.

In both cases, the gradient of state I can be explained by observing \eqref{eqn_modIepiso}. The gradient will become negative when $\hat{\gamma}I(t)+\hat{p}I(t) > \beta I(t) S(t)/N$. The difference between vaccination and non-vaccination mainly effects the $S(t)$ term, thereby causing the effect described earlier. We focus on the constant vaccination rate which decreases $S(t)$ gradually but constantly at about 0.138\% per day. At the end of the third quarter ($t=270$), we will have achieved approximately 25\% vaccinated and received immunity excluding those that have received immunity from recovery.

For overall Malaysia cases with $R_0=1.2$ (see Fig. \ref{fig_malaysia12}), we observe that peak for non-vaccination case is approximately at 1.7\% of the total population while for vaccination case, the peak is around 0.7\% of the population. The difference is significant at 0.1\%. This translates to 3.2 $\times$ 103 individuals. However, these conditions require that the infected case numbers are highly accurate. Therefore, the effects of the vaccination demonstrated by the simulations can be accepted as approximations. The difference between the two reproduction numbers produces peaks that are approximately twofold for overall Malaysia cases (compare Figs. \ref{fig_malaysia11} and \ref{fig_malaysia12}). This shows that vaccination program is effective despite the seemingly low vaccination rate of 0.138\% per day. Cumulatively, the effect increases as the susceptible population receives immunity and moves to the removed class.

The effective daily cases simulation is shown in Fig. \ref{fig_dailyprojection}. For overall Malaysia, in particular, it can be seen from Figs. \ref{fig_dailymalaysia11} and \ref{fig_dailymalaysia12} that with vaccination, both plots show peak at about 6,000 and 20,000 cases, respectively. Therefore, it is important to comply with the standard operating procedure issued by the World Health Organization (such as social distancing) to further reduce $R_0$. Similarly, plots for individual city or state show that peaks of daily cases are mostly exceeding twofold. This shows that social distancing is still much relevant and important in the fight to eradicate the pandemic while the population gets vaccinated.

    \begin{figure}[H]
	\begin{subfigure}{0.5\textwidth}
	\includegraphics[width=\linewidth]{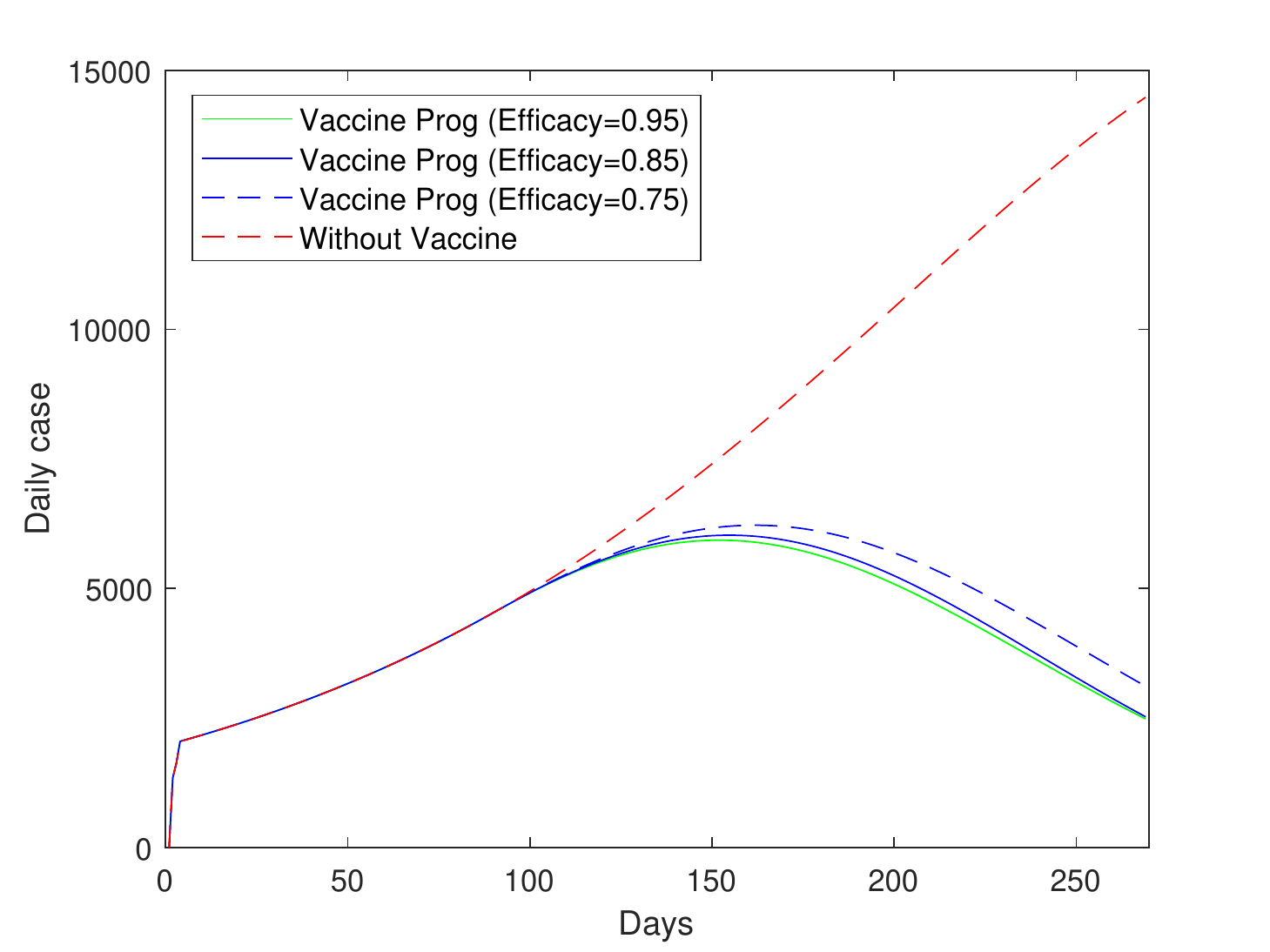}
	\caption{Malaysia, $R_0=1.1$.}
	\label{fig_dailymalaysia11}
	\end{subfigure}
	\begin{subfigure}{0.5\textwidth}
	\includegraphics[width=\linewidth]{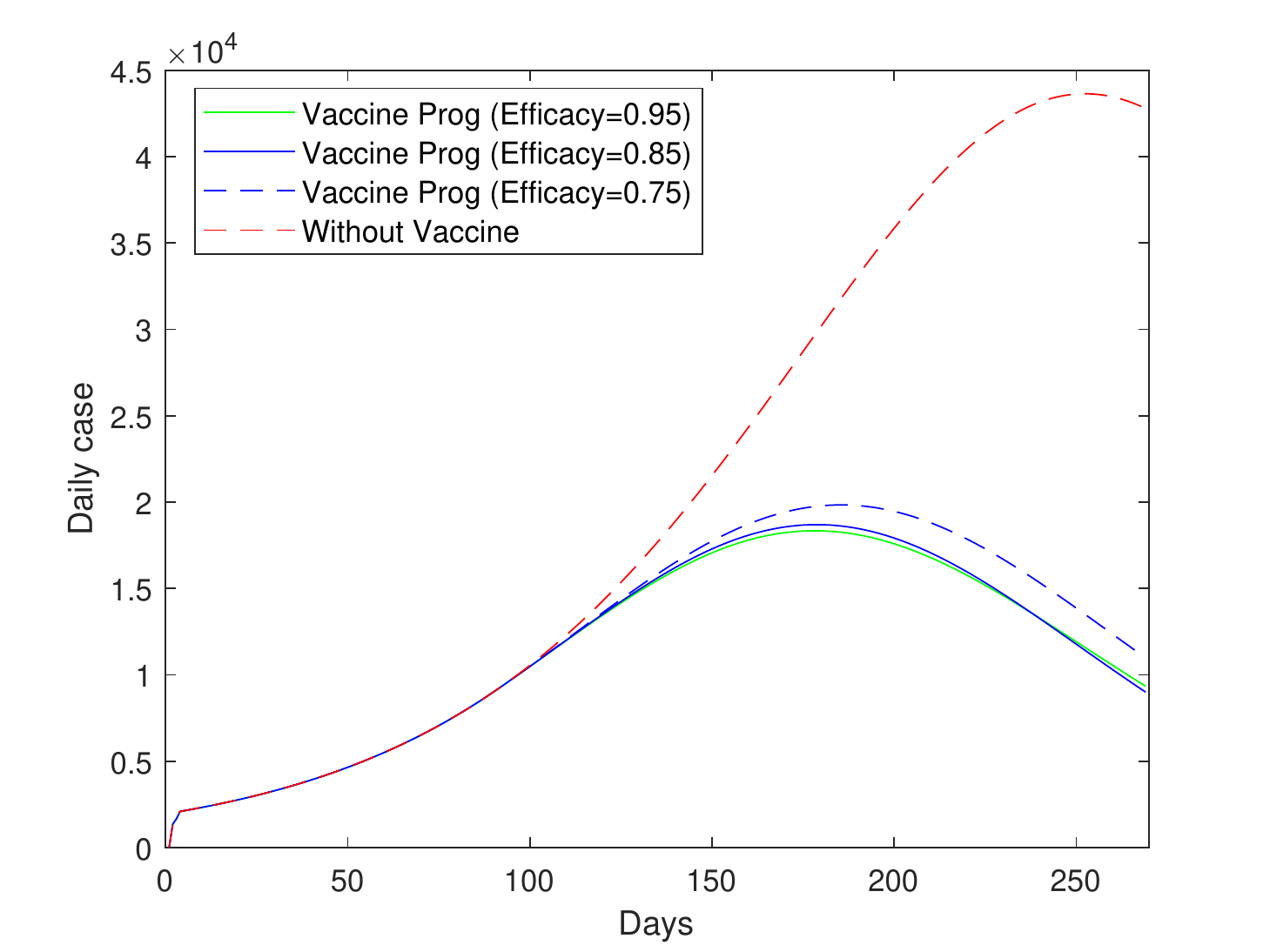}
	\caption{Malaysia, $R_0=1.2$.}
	\label{fig_dailymalaysia12}
	\end{subfigure}
	\begin{subfigure}{0.5\textwidth}
	\includegraphics[width=\linewidth]{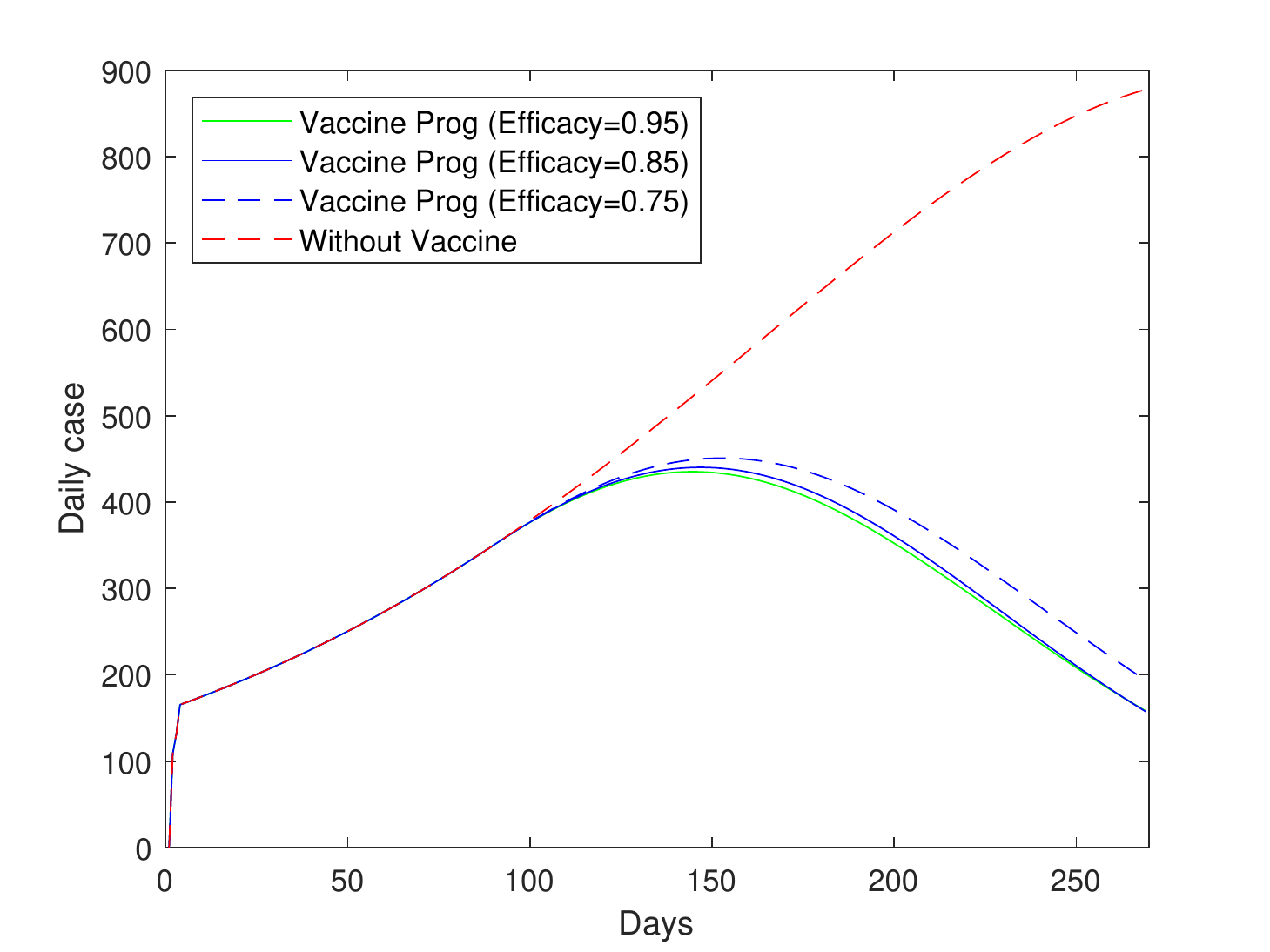}
	\caption{Kuala Lumpur, $R_0=1.1$.}
	\label{fig_kul11}
	\end{subfigure}
	\begin{subfigure}{0.5\textwidth}
	\includegraphics[width=\linewidth]{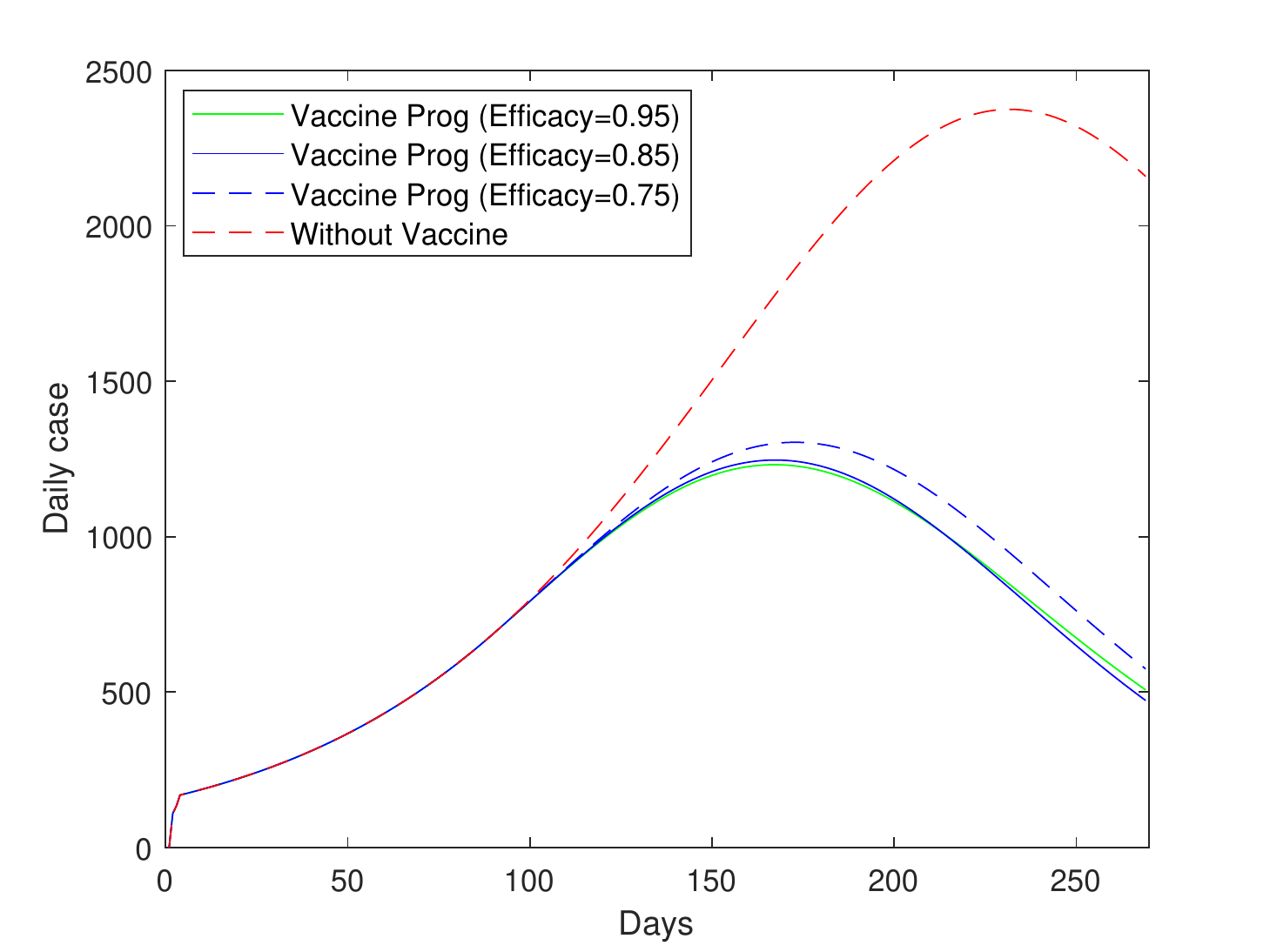}
	\caption{Kuala Lumpur, $R_0=1.2$.}
	\label{fig_kul12}
	\end{subfigure}
	\begin{subfigure}{0.5\textwidth}
	\includegraphics[width=\linewidth]{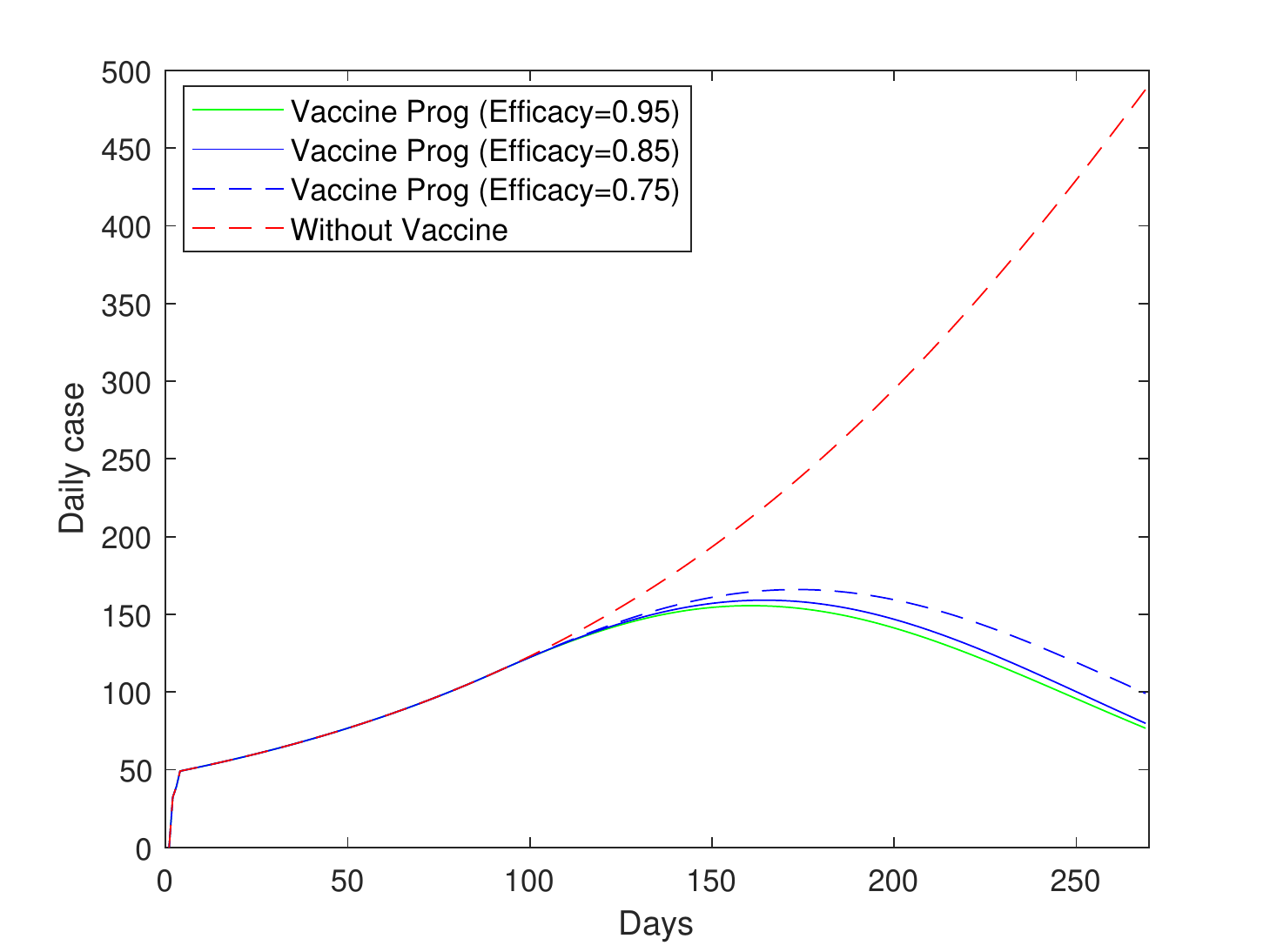}
	\caption{Penang, $R_0=1.1$.}
	\label{fig_penang11}
	\end{subfigure}
	\begin{subfigure}{0.5\textwidth}
	\includegraphics[width=\linewidth]{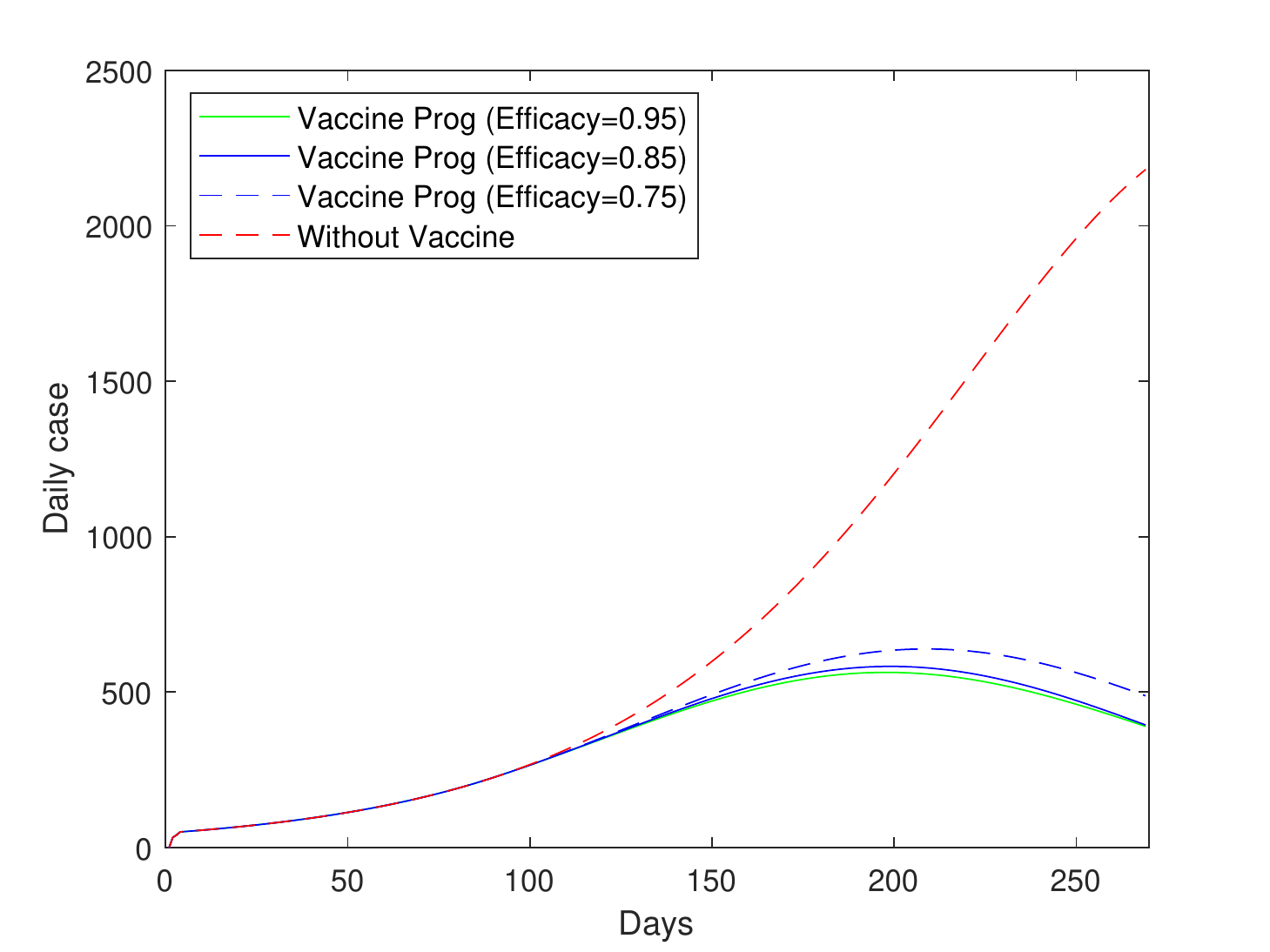}
	\caption{Penang, $R_0=1.2$.}
	\label{fig_penang12}
	\end{subfigure}
	\caption{Daily cases simulation with and without vaccination for the various localities - continued.}
	\end{figure}
	\begin{figure}[tb]\ContinuedFloat
	\begin{subfigure}{0.5\textwidth}
	\includegraphics[width=\linewidth]{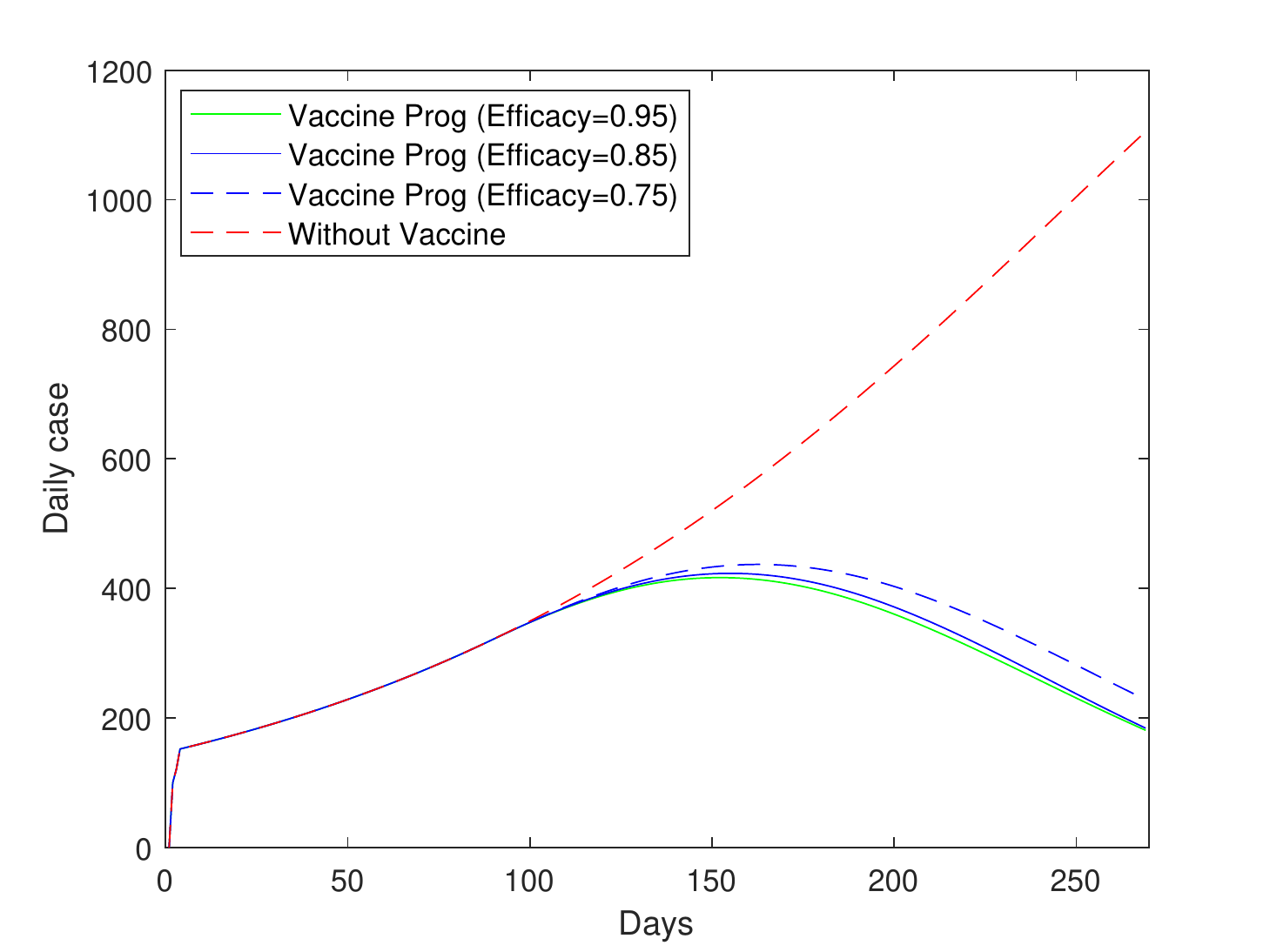}
	\caption{Sabah, $R_0=1.1$.}
	\label{fig_sabah11}
	\end{subfigure}
	\begin{subfigure}{0.5\textwidth}
	\includegraphics[width=\linewidth]{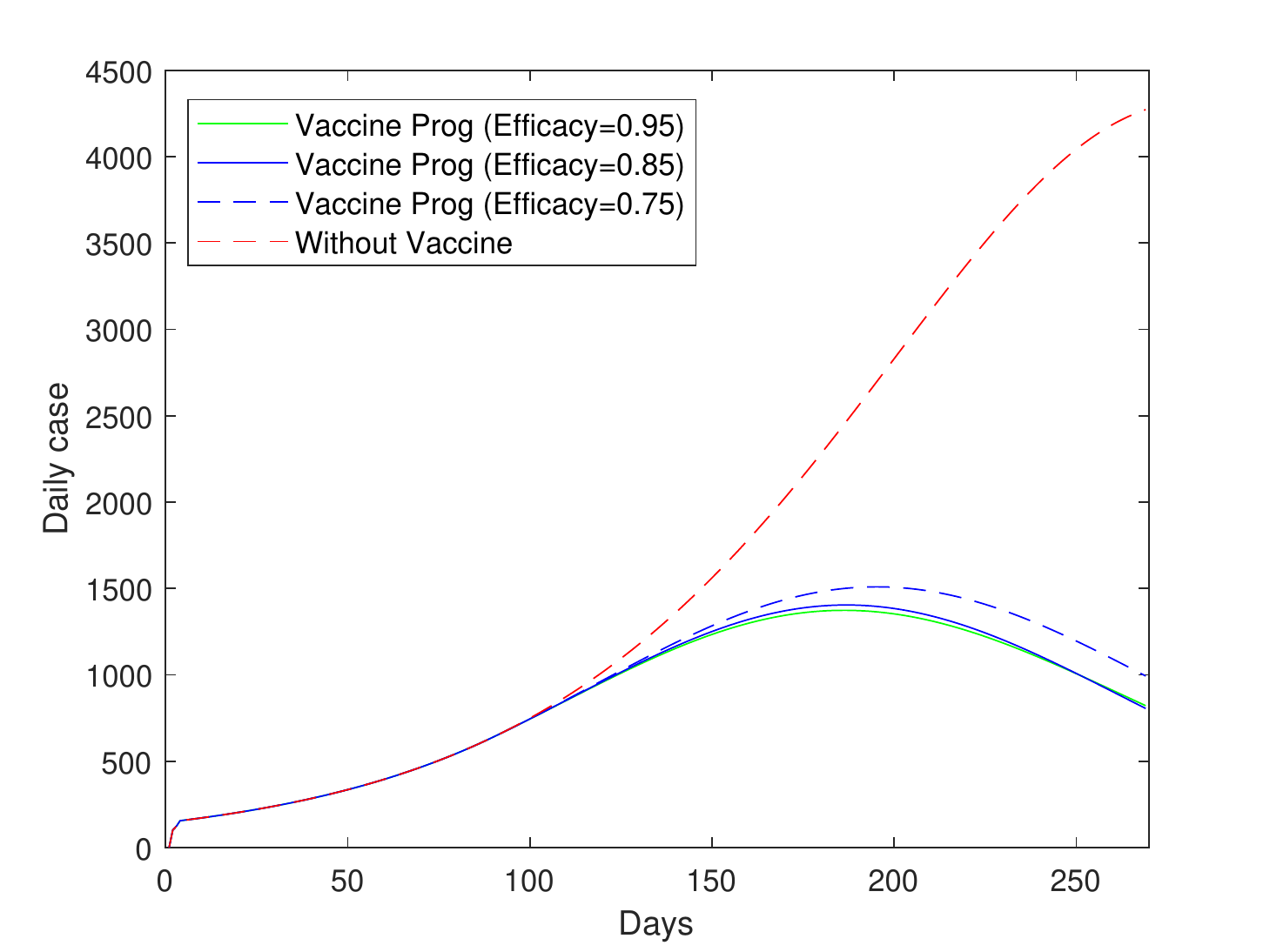}
	\caption{Sabah, $R_0=1.2$.}
	\label{fig_sabah12}
	\end{subfigure}
	\begin{subfigure}{0.5\textwidth} 
	\includegraphics[width=\linewidth]{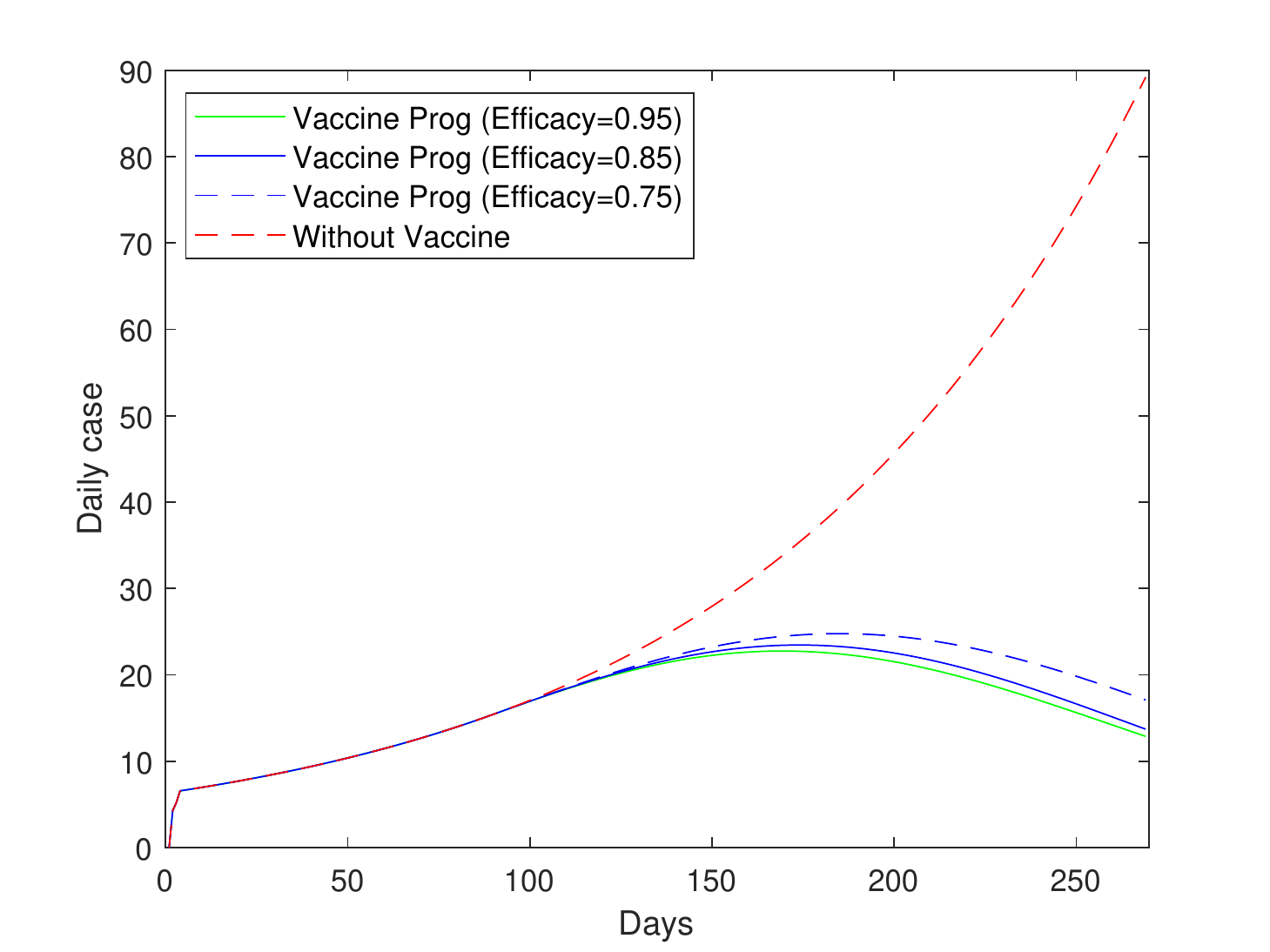}
	\caption{Sarawak, $R_0=1.1$.}
	\label{fig_sarawak11}
	\end{subfigure}
	\begin{subfigure}{0.5\textwidth}
	\includegraphics[width=\linewidth]{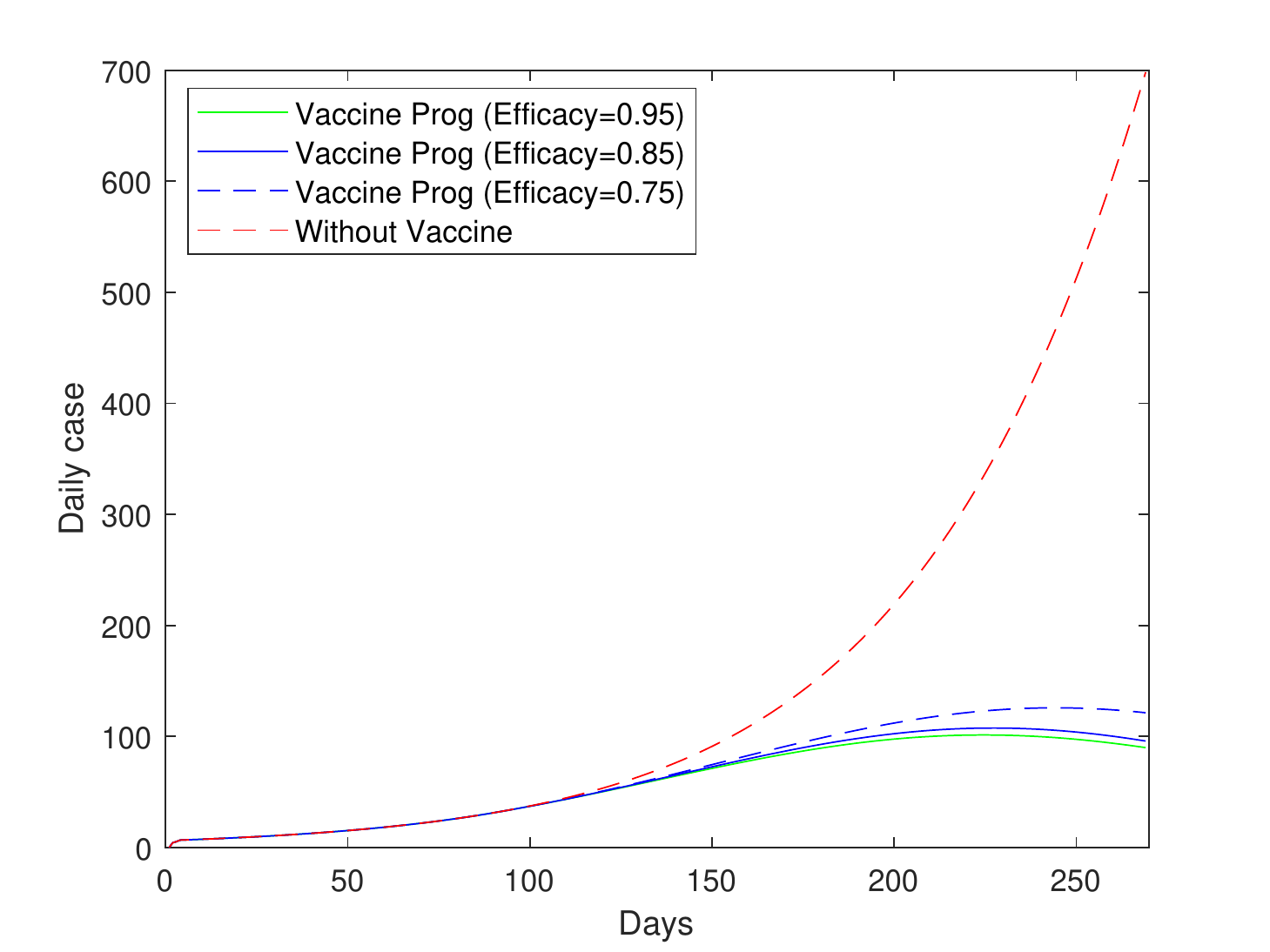}
	\caption{Sarawak, $R_0=1.2$.}
	\label{fig_sarawak12}
	\end{subfigure}
	\caption{Daily cases simulation with and without vaccination for the various localities.}
	\label{fig_dailyprojection}
	\end{figure}

\section{Conclusion}
In this paper, we have used a modified SIR model to simulate the conditions with and without the vaccination program, and at various vaccine efficacy levels.. The model applies minimal modification to the basic SIR model and hence, tuning can be easily achieved. In particular, we modify the existing SIR model to include a gradual vaccination which effectively moves population from susceptible state to removed state. In the simulation, we assume that the vaccination program will be gradual and progressive. The results have shown that vaccination makes a significant difference in combating the pandemic. With the global vaccination, we can for see that the imported cases can even be reduced. In particular, we have presented a few scenarios to predict infectious population and the daily cases in Malaysia. 

In worse case scenario, e.g. $R_0=1.2$, simulation results have shown about 20,000 daily cases daily (with vaccination). Hence, lowering $R_0$ is still very relevant and should be done simultaneously with on-going vaccination program. Based on this simulation, it is clear with the given description and assumptions, there is huge benefit in the roll out of vaccination program in Malaysia. Nevertheless, challenges may arise as the community may not want to be vaccinated, since a third of Malaysians surveyed are not convinced of the safety of COVID-19 vaccines. Therefore, community leaders should play important role in educating the public. 

Secondly, Malaysia has population of 32.7 million staying in rural areas where access can be very challenging due to infrastructure problem. A lower vaccination rate due to these circumstances may cause higher infection rates as indicated in the simulation projections. Finally, similar methodology may be extended to other countries to estimate the efficacy of vaccination on the basis of their individual data.

\bibliographystyle{elsarticle-num}
\bibliography{ref}

\end{document}